\pdfoutput=1 
\documentclass[camera,letterpaper,nomarginnotes,nonarrowgutter]{jpaper}

\usepackage[square,numbers,sort&compress]{natbib}

\usepackage{breakcites}

\usepackage{url}

\usepackage{setspace}
\usepackage[hang]{footmisc}

\usepackage{pdfpages}

\usepackage{algorithm}
\usepackage[noend]{algpseudocode}

\usepackage[colorlinks = true,
            linkcolor = blue,
            urlcolor  = blue,
            citecolor = blue,
            anchorcolor = blue,
            breaklinks
            ]{hyperref}
\usepackage{xcolor}
\definecolor{darkblue}{rgb}{0,0,0.8}

\newcommand{\secref}[1]{\hyperref[sec:#1]{\S\ref*{sec:#1}}}

\renewcommand{\algref}[1]{{\hyperref[alg:#1]{Alg.~\ref*{alg:#1}}}}

\newcommand{\alglineref}[1]{{\hyperref[algline:#1]{Line~\ref*{algline:#1}}}}

\newcommand{\algrangeref}[2]{{\hyperref[algline:#1]{Lines~\ref*{algline:#1}\nobreakdash-\ref*{algline:#2}}}}

\newcommand{\figref}[2][]{{\hyperref[fig:#2]{Fig.~\ref*{fig:#2}#1}}}

\newcommand{\tabref}[1]{{\hyperref[table:#1]{Tab.~\ref*{table:#1}}}}

\newcommand{\theoref}[1]{{\hyperref[theorem:#1]{Theorem~\ref*{theorem:#1}}}}

\usepackage{xspace}
\newcommand{\toolname}[0]{Scrooge\xspace}
\newcommand{\gitrepo}[0]{\url{https://github.com/cmu-safari/Scrooge}}

\newcounter{insight}

\newcommand{\insightref}[1]{{\hyperref[insight:#1]{Observation~\ref*{insight:#1}}}}

\newcommand{\cpuoptspeedup}[0]{2.1\x}
\newcommand{\kswcpuspeedup}[0]{20.1\x}
\newcommand{\edlibcpuspeedup}[0]{1.7\x}

\newcommand{\cpugpuspeedup}[0]{4.0\x}
\newcommand{\gpuoptspeedup}[0]{5.9\x}
\newcommand{\kswgpuspeedup}[0]{80.4\x}
\newcommand{\edlibgpuspeedup}[0]{6.8\x}
\newcommand{\darwingpuspeedup}[0]{12.6\x}

\newcommand{\defn}[1]{\textit{#1}}
\newcommand{\x}[0]{$\times$\xspace}
\newcommand{\var}[1]{\texttt{#1}}

\usepackage{amsmath}

\DeclareMathOperator*{\argmin}{arg\,min}

\newcommand*\BitAnd{\mathbin{\&}}
\newcommand*\BitOr{\mathbin{|}}
\newcommand*\ShiftLeft{\ll}

\definecolor{lightorange}{rgb}{1.0,0.8,0.6}
\usepackage[colorinlistoftodos,prependcaption,textsize=scriptsize,color=lightorange]{todonotes}
\newtheorem{theorem}{Theorem}

\newcommand{\bitapfigwidth}[0]{0.6}

\usepackage{fancyhdr}
\usepackage{balance}

\title{Scrooge: A Fast and Memory-Frugal Genomic Sequence Aligner\\for CPUs, GPUs, and ASICs}
\newcommand{\affilETH}[0]{\textsuperscript{\S}}
\newcommand{\affilBionano}[0]{\textsuperscript{$\dagger$}}

\newcommand{\authorgap}[0]{\hspace{6em}}
\author{
{Joël Lindegger\affilETH}\authorgap%
{Damla Senol Cali\affilBionano}\authorgap%
{Mohammed Alser\affilETH}\\%
\hspace{-1.5em}{Juan~Gómez-Luna\affilETH}\authorgap\hspace{-1.5em}%
{Nika Mansouri Ghiasi\affilETH}\hspace{0.5em}\authorgap%
{Onur Mutlu\affilETH}
\vspace{2mm}\\%
\emph{\affilETH ETH Z{\"u}rich~~~~~~~~~ \affilBionano Bionano Genomics} 
}

\begin{document}
\bstctlcite{IEEEexample:BSTcontrol}

\maketitle
\thispagestyle{plain}

\begin{abstract}Pairwise sequence alignment is a  very time-consuming step in common bioinformatics pipelines. Speeding up this step requires heuristics, efficient implementations, and/or hardware acceleration. A promising candidate for all of the above is the recently proposed GenASM algorithm. We identify and address three inefficiencies in the GenASM algorithm: it has a high amount of data movement, a large memory footprint, and does some unnecessary work.

We propose \defn\toolname, a fast and memory-frugal genomic sequence aligner. \toolname includes three novel algorithmic improvements which reduce the  data movement, memory footprint, and the number of operations in the GenASM algorithm. We provide efficient open-source implementations of the \toolname algorithm for CPUs and GPUs, which demonstrate the significant benefits of our algorithmic improvements.
For long reads, the CPU version of \toolname achieves a \kswcpuspeedup, \edlibcpuspeedup, and \cpuoptspeedup speedup over KSW2, Edlib, and a CPU implementation of GenASM, respectively. The GPU version of \toolname achieves a \cpugpuspeedup \kswgpuspeedup, \edlibgpuspeedup, \darwingpuspeedup and \gpuoptspeedup speedup over the CPU version of \toolname, KSW2, Edlib, Darwin-GPU, and a GPU implementation of GenASM, respectively. We estimate an ASIC implementation of \toolname to use 3.6\x less chip area and 2.1\x less power than a GenASM ASIC while maintaining the same throughput. Further, we systematically analyze the throughput and accuracy behavior of GenASM and \toolname under various configurations. As the best configuration of \toolname depends on the computing platform, we make several observations that can help guide future implementations of \toolname.\\
\textbf{Availability:} \url{https://github.com/CMU-SAFARI/Scrooge}
\end{abstract}

\section{Introduction} \label{sec:introduction}
\defn{Pairwise sequence alignment} is a computational step commonly required in bioinformatics pipelines~\citep{alser2022frommolecules}, such as in \defn{read mapping}~\cite{alser2020accelerating} and \defn{de-novo assembly}~\citep{li2011denovo}. We formulate the problem as (1)~finding the \defn{edit distance} between two sequences~\citep{levenshtein1966binary} and (2)~determining the sequence of corresponding edits. Efficient algorithms for solving this problem optimally are based on \defn{dynamic programming (DP)}, such as the Smith-Waterman-Gotoh algorithm~\citep{smith1981identification,gotoh1982improved}, and have a runtime that grows quadratically with sequence length~\citep{alser2020technology}. \citep{asm_lower_bound}~proves no strongly subquadratic time solutions can exist, provided the strong exponential time hypothesis~\citep{impagliazzo2001complexity} holds. Hence, recent works focus on approaches such as pre-alignment filtering~(e.g.,~\citep{xin2013fasthash,xin2015shiftedhammingdistance,alser2019shouji,alser2020sneakysnake,singh2021fpga,mansouri2022genstore}), constant factor algorithmic speedups~(e.g.,~\citep{edlib, ksw2_a, ksw2_b, wfa_algorithm}), GPU-based acceleration (e.g., \citep{gasal2,ahmed2020darwin_gpu,de2016cudalign,liu2013cudasw++,awan2020adept}), FPGA-based acceleration~(e.g.,~\citep{fei2018fpgasw, hoffmann2016using, benkrid2009highly}) or using specialized hardware accelerators~(e.g., \citep{ turakhia2019darwin,turakhia2019darwinwga,fujiki2018genax,fujiki2020seedex,cali2020genasm,cali2022segram}).

We observe that GenASM~\citep{cali2020genasm}, a recent state-of-the-art sequence alignment algorithm, has a large space for improvement. %
GenASM uses only cheap bitwise operations and breaks the lower complexity bound of pairwise sequence alignment through its powerful \defn{windowing heuristic}. \citep{cali2020genasm} has already proven the effectiveness of the GenASM algorithm and its accelerator implementation, thus we are motivated to further improve the GenASM algorithm and explore its potential on commodity hardware.

We identify \textbf{three inefficiencies} in the GenASM algorithm: (1)~it has a \emph{large memory footprint} due to the large size of the dynamic programming (DP) table, (2)~it has a \emph{high amount of data movement} between registers and memory due to frequent accesses to the DP table, and (3)~it does some \emph{unnecessary} work by calculating DP cells that are not useful for finding the final result.
The three inefficiencies negatively impact both (1)~software implementations running on commodity hardware (e.g., CPUs or GPUs) and (2)~custom hardware (e.g., ASIC) implementations.

\emph{Software implementations} on commodity hardware typically cannot
fit all the data into fast on-chip memories (e.g., L1, scratchpad memory) due to the large memory footprint. This increases the latency and limits the bandwidth with which the DP table can be accessed. The high amount of data movement puts high pressure on this bandwidth, limiting performance.

In contrast, \emph{custom hardware implementations} can use arbitrarily large amounts of on-chip memory, but such a large on-chip memory with the high bandwidth requirement is costly. For example, the hardware accelerator described in~\citep{cali2020genasm} requires 76\% and 54\% of the total chip area and power consumption
for the on-chip memory that stores the DP table.

The unnecessary work stems from computing cells that do not contain useful information for finding the final result. This applies to at least 25\% of cells on average for uncorrelated string pairs, and more for correlated string pairs, as we show in \secref{early_termination}. Doing unnecessary work affects software and hardware implementations equally because both could use the wasted time to do useful work instead.

Our \textbf{goal}
is to develop a fast and memory-frugal alignment algorithm by addressing the inefficiencies in the GenASM algorithm, and demonstrate its benefits with high-performance CPU and GPU implementations.

To this end we propose \defn\toolname\footnote{\emph{\toolname} is a name for miserly or frugal fictional characters (e.g.,~\citep{dickens1843achristmascarol}), similar to how our proposed algorithm aims to be as resource-efficient as possible.}, which includes improvements to the GenASM algorithm based on three \textbf{key~ideas}:
\begin{enumerate}
    \item The DP table can be \emph{compressed} by storing only the bitwise AND of multiple values (\secref{SENE}). The required regions of the DP table can then be decompressed on-demand during traceback with a small computational overhead.
    \item Part of the DP table \emph{does not need to be stored} because the traceback operation cannot reach these entries (\secref{DENT}).
    \item Part of the DP table can opportunistically be \emph{excluded from calculation} if previous rows of the DP table already contain the information needed for finding the final result (\secref{early_termination}).
\end{enumerate}
These improvements (1)~reduce the number of accesses to GenASM's DP table, (2)~reduce the memory footprint of the DP table, and (3)~eliminate unnecessary work. %

We experimentally demonstrate that our improvements yield significant benefits across multiple computing platforms and multiple baseline sequence alignment methods. The CPU version of \toolname achieves a \kswcpuspeedup, \edlibcpuspeedup, and \cpuoptspeedup speedup over CPU-based implementations of KSW2~\citep{ksw2_a,ksw2_b}, Edlib~\citep{edlib}, GenASM, respectively. The GPU version of \toolname achieves a \cpugpuspeedup, \kswgpuspeedup, \edlibgpuspeedup, \darwingpuspeedup, and \gpuoptspeedup speedup over CPU-based implementations of \toolname, KSW2, and Edlib, and GPU-based implementations of Darwin-GPU~\citep{ahmed2020darwin_gpu} and GenASM, respectively. We analytically estimate an ASIC implementation of \toolname to use 3.6\x less chip area and consume 2.1\x less power compared to the prior state-of-the-art ASIC implementation of GenASM~\citep{cali2020genasm} while maintaining the same throughput.

The \textbf{contributions} of this paper are as follows:
\begin{itemize}
    \item We develop three novel algorithmic improvements that are applicable to software and custom hardware implementations of \toolname, collectively reducing the memory footprint by 24\x, the number of memory accesses by 12\x, and the number of entries of the DP table calculated by at least 25\% on average compared to GenASM.
    \item We experimentally demonstrate the significant throughput (i.e., alignments per second) increase of our improvements for CPU and GPU implementations of \toolname.
    \item We analytically estimate that an ASIC implementation of \toolname\ significantly reduces the chip area and power consumption compared to the prior state-of-the-art ASIC implementation of GenASM.
    \item We open-source all code, including high-performance CPU and GPU implementations of \toolname, which can be readily used as a sequence alignment library, and all evaluation scripts.
    \item We systematically analyze the throughput and accuracy behavior of GenASM and \toolname across a range of configurations based on real and simulated datasets for long and short reads. As the best configuration of \toolname depends on the computing platform, we make several observations that can help guide future implementations of \toolname.
\end{itemize}

\section{Materials and methods} \label{sec:methods}
\subsection{Overview} \label{sec:overview}
The primary purpose of \toolname is to accelerate pairwise sequence alignment through (1)~a memory-frugal and efficient algorithm, and (2)~optimized CPU, GPU, and ASIC implementations.

\toolname solves the \defn{approximate string matching (ASM)} problem with the \defn{edit distance}~\citep{levenshtein1966binary} as the cost metric. That is, given two strings, \var{text} and \var{pattern}, \toolname finds the minimum number of single-letter substitutions, insertions, and deletions to convert \var{text} into \var{pattern}. Additionally, the sequence of edits that corresponds the edit distance is reported, which is called \defn{CIGAR string}.

The \toolname algorithm is based on the GenASM algorithm~(\secref{genasm_algo}).
\citep{cali2020genasm} first proposed the GenASM algorithm as an algorithm/hardware co-design targeted for an ASIC accelerator, and demonstrated GenASM's potential for very high throughput and resource efficiency. However, as we show in \secref{challenges}, the GenASM algorithm (1)~requires large amounts of memory bandwidth, (2)~exhibits a large memory footprint, and (3)~does some unnecessary work. These inefficiencies limit GenASM's throughput and resource efficiency on both commodity and custom hardware, and addressing them is critical.

To this end, we propose \toolname's three novel algorithmic improvements to GenASM in \secref{optimizations}. 
In \secref{throughput}, we experimentally demonstrate that these improvements significantly increase performance on recent CPUs and GPUs. In \secref{sensitivity}, we explore the throughput behavior of GenASM with and without the proposed improvements across various configurations. We show in \secref{hw_benefits} that an ASIC implementation of \toolname will have significantly reduced chip area and power consumption compared to the ASIC designed for GenASM~\citep{cali2020genasm} while maintaining the same throughput. In \secref{accuracy}, we explore the accuracy behavior of GenASM and \toolname across various configurations.

\subsection{GenASM Algorithm} \label{sec:genasm_algo}
The GenASM algorithm~\citep{cali2020genasm} consists of two sub-algorithms: \emph{GenASM-DC} and \emph{GenASM-TB}. GenASM-DC~(\secref{bitap}) fills a bitvector-based dynamic programming table. The last column of the table indicates the edit distance between the two input strings. GenASM-TB (\secref{traceback}) re-traces this optimal solution in the constructed table. To better scale with longer input sequences, GenASM uses a \emph{windowing heuristic} (\secref{windowing}).

\subsubsection{GenASM-DC Algorithm} \label{sec:bitap}
\begin{figure}[h]
    \centering
    \includegraphics[trim={8mm 0 0 0},width=0.87\columnwidth,keepaspectratio]{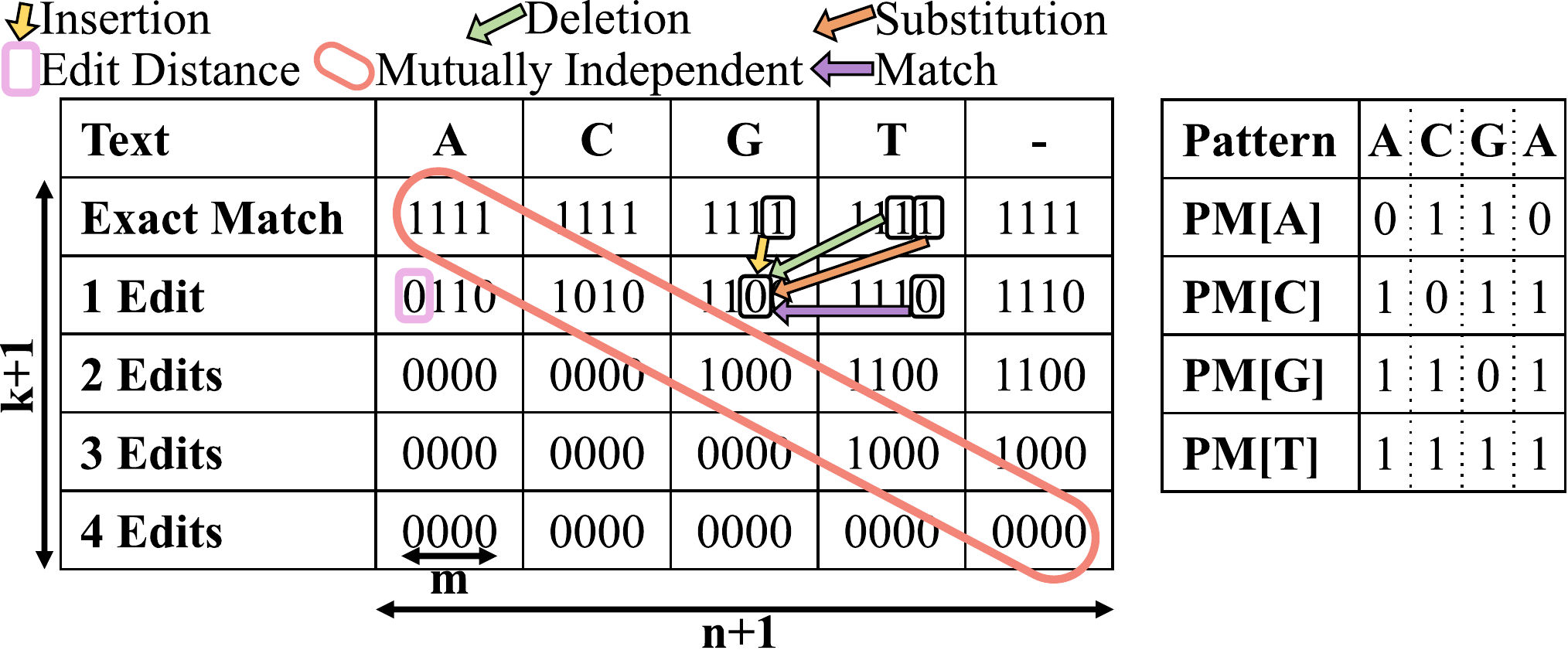}
    \caption{An example of DP table \var{R} with \var{text}=\texttt{ACGT} and \var{k}=4. The bitmasks for \var{pattern}=\texttt{ACGA} are shown on the right. The colored arrows show the possible origins and data dependencies of the $0$ at \var{d}=1,\var{i}=2,\var{j}=2. The values in the red marked diagonal are mutually independent and thus can be computed in parallel.}
    \label{fig:bitap_example}
\end{figure}

GenASM-DC uses only cheap bitwise operations to calculate the edit distance between two strings \var{text} and \var{pattern}~\citep{cali2020genasm}.
It builds an (\var{n}+1)$\times$(\var{k}+1) dynamic programming (DP) table \var{R}, where \var{n}=$length$(\var{text}) and \var{k} is the maximum number of edits considered. The entries of \var{R} are \var{m}-bit bitvectors, where \var{m}=$length$(\var{pattern}). \figref{bitap_example} shows an example of \var{R} after it is constructed by GenASM-DC.

\begin{theorem}
    \label{theorem:R_interpretation}
    The entries (bitvectors) of \var{R} can be interpreted as follows:
    \vspace{-5px}\begin{align*}
    \var{j}\text{-th bit of }\var{R}[\var{i}][\var{d}] = 0 \Longleftrightarrow \\[-5pt]
    distance\big(\var{text}[\var{i}:\var{n}), \var{pattern}[\var{j}:\var{m})\big) \leq \var{d}
    \end{align*}
\end{theorem}

In natural language, \theoref{R_interpretation} states that the \var{j}-th bit of the bitvector \var{R[i][d]} is 0 exactly if the suffix of \var{text} starting at character \var{i} and the suffix of \var{pattern} starting at character \var{j} differ by at most \var{d} edits. Following this interpretation, the first row \var{d=d$_{OPT}$} that has a \var{0} in the first bit (\var{j=0}) of the leftmost column (\var{i=0}) indicates that the edit distance between \var{text} and \var{pattern} is \var{d$_{OPT}$}.  This bit is marked in pink in~\figref{bitap_example}.

GenASM-DC~(\algref{bitap}) starts by pre-processing \var{pattern} into four \defn{pattern masks}, one per character in the alphabet. The pattern mask for character \var{X}{$\in$}$\{A,C,G,T\}$ is a bitvector of length \var{m}=$length$(\var{pattern}), with a 0 in the \var{i}-th bit if \var{pattern}[\var{i}]==\var{X}. See \figref{bitap_example} for an example.

GenASM-DC populates the rightmost column (\alglineref{right_init}) and topmost row (\alglineref{top_init}) of \var{R}. The remaining entries are then calculated from their respective neighbors in the north (\alglineref{insertion}, insertion), north-east (\algrangeref{deletion}{substitution}, deletion and substitution), and east (\alglineref{match}, match) through simple bitwise update rules. We refer to ~\citep{cali2020genasm,wu_manber_bitap,baeza-yates-gonnet-1992} for detailed arguments on the correctness of GenASM-DC.
To follow the rest of this paper, it is sufficient to consider (1)~the interpretation of \var{R} given in \theoref{R_interpretation}, and (2)~the north-east data dependencies imposed by \algref{bitap} and shown in \figref{bitap_example}.

\begin{algorithm}
\begin{spacing}{0.78}
\caption{GenASM-DC Algorithm}\label{alg:bitap}
\textbf{Inputs:} \var{text}, \var{pattern}, \var{k} \\
\textbf{Outputs:} \var{editDist}

\begin{algorithmic}[1]
    \State $\var{n} \gets \Call{length}{\var{text}}$
    \State $\var{m} \gets \Call{length}{\var{pattern}}$
    \State $\var{PM} \gets \Call{buildPatternMasks}{\var{pattern}}$
    \State
    \State $\var{R}[\var{n}][\var{d}] \gets 11...1 \ShiftLeft \var{d}$ \label{algline:right_init}
    \Comment{Initialize for all $0 \leq \var{d} \leq \var{k}$ }
    \State
    \For{\var{i} \textbf{in} $(\var{n}-1):-1:0$}
        \State $\var{char} \gets \var{text}[\var{i}]$
        \State $\var{curPM} \gets \var{PM}[\var{char}]$
        \State
        \State $\var{R}[\var{i}][0] \gets (\var{R}[\var{i}+1][0] \ShiftLeft 1) \BitOr \var{curPM}$ \label{algline:top_init}
        \Comment{exact match}
        \For{\var{d} \textbf{in} $1:\var{k}$}
            \State $\var{I} \gets \var{R}[\var{i}][\var{d}-1] \ShiftLeft 1$ \label{algline:update_rules_start}\label{algline:insertion}
            \Comment{insertion}
            \State $\var{D} \gets \var{R}[\var{i}+1][\var{d}-1]$\label{algline:deletion}
            \Comment{deletion}
            \State{$\var{S} \gets \var{R}[\var{i}+1][\var{d}-1] \ShiftLeft 1$}\label{algline:substitution}
            \Comment{substitution}
            \State{$\var{M} \gets (\var{R}[\var{i}+1][\var{d}] \ShiftLeft 1) \BitOr \var{curPM}$}\label{algline:match}
            \Comment{match}
            \State $\var{R}[\var{i}][\var{d}] \gets \var{I} \BitAnd \var{D} \BitAnd \var{S} \BitAnd \var{M}$ \label{algline:update_rules_end}
        \EndFor
    \EndFor
    \State
    \State $\var{editDist} \gets \argmin_\var{d}\{\Call{msb}{\var{R}[0][\var{d}]} = 0\}$ \label{algline:edit_dist}
\end{algorithmic}
\end{spacing}
\end{algorithm}

\noindent\textbf{Intra-Task Parallelism.} \label{sec:task_parallelism}
\citep{cali2020genasm} enables efficient intra-task parallelism by identifying that the DP entries within each north-west to south-east diagonal (one such diagonal is marked in red in \figref{bitap_example}) do \emph{not} depend on each other, hence they can be computed in parallel.

\subsubsection{GenASM-TB Algorithm} \label{sec:traceback}
For use-cases like read mapping, the pairwise sequence alignment algorithm should report both the edit distance and the corresponding sequence of edits, which is called the \defn{CIGAR string}. Obtaining the CIGAR string involves retracing the origin of the edit distance value as a linear path through DP entries in their reverse construction order; this process is called \defn{traceback}.

GenASM enables efficient traceback operations based on two key observations:
First, if \emph{all} intermediate values of variables \var{I}, \var{D}, \var{S} and \var{M} in \algref{bitap} are stored, then one can follow the path of \var{0}s in these variables, starting from \var{0} in the west of \var{R} that indicates the edit distance (highlighted in pink in \figref{bitap_example}) and go towards the north-east corner of \var{R}. Whenever a 0 in one of these variables is traversed, the name of that variable is recorded as an edit (e.g., '\var{I}' for an insertion). Second, it is sufficient to store only three out of the four variables (because \var{S} can be obtained by shifting \var{D}), saving both memory footprint and bandwidth. %

\subsubsection{GenASM's Windowing Heuristic} \label{sec:windowing}

To provide a linear runtime complexity, \citep{cali2020genasm} proposes a greedy \defn{windowing heuristic}. Instead of aligning \var{text} and \var{pattern} in a single run of GenASM-DC, the windowing heuristic runs GenASM-DC multiple times as a subroutine in \defn{windows} of size \var{W}. In each window, a prefix of size \var{W} characters of each sequence (i.e., $\var{text}[0:\var{W})$ and $\var{pattern}[0:\var{W})$) are aligned. The first $\var{W} - \var{O}$ characters of the window are greedily considered aligned optimally, where we call \var{O} the window \defn{overlap}. The smaller strings $\var{text}[\var{W}-\var{O}:\var{n})$ and $\var{pattern}[\var{W}-\var{O}:\var{m})$ then remain to be aligned in the next window.

This approach has three advantages. First, instead of constructing a large table of \var{n}$\times$\var{m}$\times$\var{k} bits, only $\frac{\var{m}}{\var{W} - \var{O}}$ tables of $\var{W}^3$ bits must be constructed, saving memory footprint, data movement, and computation. Second, the GenASM-DC subroutine now runs over constant-sized sequences, simplifying its implementation. For example, DP entries can be statically assigned to processing elements~\citep{cali2020genasm}, and the data movement and exact memory footprint are known at compile time, even if the length of the input sequences is unknown. Third, the program flow (e.g., the number of loop iterations per window) is entirely known at compile time, giving the compiler the ability to optimize.

The windowing strategy is greedy and heuristic, so it is possible that it could miss the optimal alignment and produce a sub-optimal one instead. This is a key limitation of GenASM and \toolname. Note that several state-of-the-art tools do not give any optimality guarantees either, and instead experimentally demonstrate their practical accuracy, as \toolname does. This includes greedy alignment techniques like SeGraM~\citep{cali2022segram}, Darwin~\citep{turakhia2019darwin}, and WFA-adaptive~\citep{wfa_algorithm}, as well as mappers based on sparse dynamic programming, like minimap2~\citep{ksw2_b}. To balance performance and accuracy, the tunable parameters \var{W}~(window size) and \var{O}~(window overlap) must be selected appropriately.
The parameter \var{W} can be understood as the \emph{range} of solutions considered, similar to the \defn{band width}~\citep{UKKONEN1985banded} in popular alignment implementations~(e.g. \citep{edlib,ksw2_a,ksw2_b}). The parameter \var{O} can be understood as the \emph{globality} of the solutions or \emph{inverse greediness}. We demonstrate in \secref{results} (1)~that higher \var{W} and \var{O} generally improve accuracy, at the cost of lowering throughput, (2)~that the best choice of \var{W} and \var{O} depends on the input dataset (e.g., its error distribution and read lengths), and (3)~that \var{W}=64 and \var{O}=33 achieve a good throughput/accuracy tradeoff for long and short read mapping.

\subsection{Inefficiencies in the GenASM Algorithm} \label{sec:challenges}
We identify three inefficiencies in the GenASM algorithm: (1)~it has a large amount of data movement, (2)~it has a large memory
footprint, and (3)~it does some unnecessary work.

The combination of large amount of data movement and large memory footprint, which we quantify in \secref{data_movement_and_roofline_model} and~\secref{memory_footprint} respectively, affects both software implementations running on commodity hardware, as well as custom hardware implementations. 
Commodity hardware (e.g., CPUs or GPUs) has a fixed amount of on-chip memory. The DP table might not fit into this on-chip memory, which introduces three inefficiencies: Data has to be moved a larger distance, which increases (1)~access latency and (2)~access energy~\citep{boroumand2018google}. (3)~The high amount of data movement puts high pressure on memory bandwidth, which is scarce when accessing data residing off-chip. This causes the entire application to become memory bandwidth-bound, thus wasting compute resources and achieving sub-optimal performance. Custom hardware implementations (e.g. ASICs) can have as large on-chip memory as needed, but such a large and high-bandwidth on-chip memory comes at the cost of a large chip area and power consumption (\citep{boroumand2021neuraledgedevices}).

Doing unnecessary work trivially wastes runtime and energy. In \secref{early_termination}, we identify the DP entries that are calculated needlessly by GenASM, and quantify how frequent they are.

\subsubsection{Roofline Model}
\label{sec:data_movement_and_roofline_model}
We use the \defn{roofline model}~\citep{seminal_roofline, applying_roofline} to visualize that GenASM has a large amount of data movement, and that its operational intensity (i.e., the number of operations per byte) is too low to saturate the compute resources of modern CPUs and GPUs. The roofline model plots the upper limit of achievable compute throughput for different operational intensities for a given processor. It consists of horizontal peak compute throughput rooflines, and sloped memory bandwidth rooflines.

\begin{figure}[h]
    \centering
    \includegraphics[width=\columnwidth,keepaspectratio]{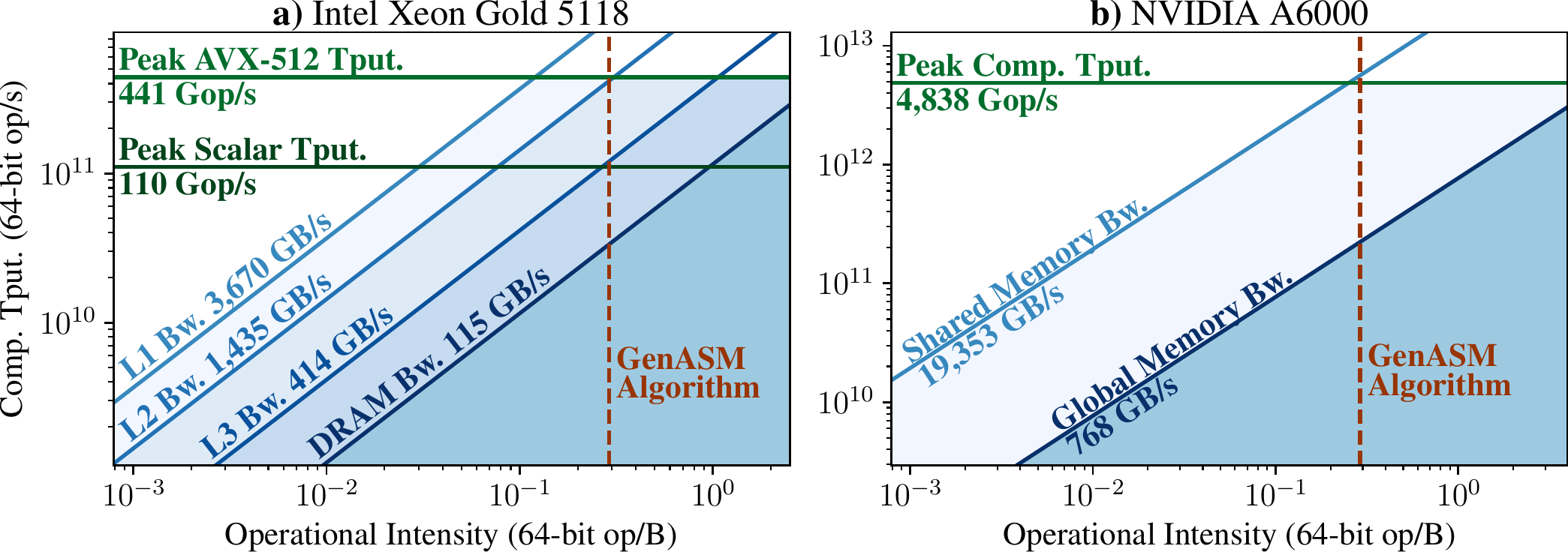}
    \caption{The roofline models of a) an Intel Xeon Gold 5118 CPU and b) an NVIDIA A6000 GPU.}
    \label{fig:roofline}
\end{figure}

\figref{roofline} shows the roofline plots for an Intel Xeon Gold 5118 CPU~\citep{xeon_gold_5118} and an NVIDIA A6000 GPU~\citep{A6000}, including their respective on-chip memory (\defn{shared memory} in CUDA), cache, and off-chip memory (\defn{global memory} in CUDA) bandwidths (drawn in shades of blue) and peak compute throughputs (draw in shades of green). GenASM's operational intensity is drawn in red. We derive the roofline parameters in \S8 of the supplementary materials.

From \figref{roofline} we make three observations. First, if the data resides off-chip, GenASM is heavily memory bandwidth-bound for a modern CPU and GPU. This is evidenced by the red (algorithm) and dark blue (off-chip memory bandwidth) lines intersecting far below the green (peak compute throughput) line.
Second, GenASM would no longer be memory bandwidth-bound if its computational intensity were $\geq$10\x higher, because then the red (algorithm) line would be shifted to the right and intersect with the dark blue (off-chip memory bandwidth) line above the green (peak compute throughput) line. The operational intensity could be increased by reducing GenASM's data movement.
Third, if the data resides in the fastest on-chip memory, GenASM \emph{can} reach peak compute throughput, even with the high amount of data movement in the baseline algorithm. This is evidenced by the red (algorithm) and light blue (L1/shared memory bandwidth) lines intersecting above the green (peak compute throughput) line. However, as we show in \secref{memory_footprint}, GenASM's memory footprint is too large for the typical capacity of such fast on-chip memories in commodity hardware, and building large enough on-chip memories is costly.

Based on these observations, we conclude that (1)~GenASM cannot saturate commodity hardware with computation, and (2)~data movement should be reduced to address this inefficiency.

\subsubsection{Memory Footprint} \label{sec:memory_footprint}
In this section, we demonstrate the overheads associated with GenASM's large memory footprint.%

We derive GenASM's working set memory footprint to be 96.5$KiB$ in \S9 of the supplementary materials. For comparison, the Intel Xeon Gold 5118 has 32$KiB$ of L1D cache per core~\citep{xeon_gold_5118} and NVIDIA's \defn{Ampere} GPU microarchitecture provides up to 99$KiB$ of high-bandwidth on-chip memory per GPU core (\defn{streaming multiprocessor}, \defn{SM} in CUDA)~\citep{cuda_guide}. Thus, one SM can hold the DP table for exactly one GenASM problem instance in its on-chip memory. One thread block of two warps (i.e., 2$\times$32 threads) can work on a single GenASM problem instance, but this does not saturate the compute resources in the SM. This is because modern GPUs are designed to alternate between executing \emph{multiple} independent instruction streams for the purpose of hiding the latency of instructions~\citep{lindholm2008nvidia}. Underutilization of the compute resources in an SM due to too few independent instruction streams is called \defn{low occupancy} and causes the unused computational resources to be wasted~\citep{cuda_guide}. Hence, the occupancy should be increased by working on multiple problem instances per SM. Multiple problem instances can fit into memory by \emph{either} reducing the memory footprint per problem instance, \emph{or} placing the DP tables into the GPU's off-chip memory, which has a much larger capacity. Our goal is the former, as we show in \secref{data_movement_and_roofline_model} that the latter is \emph{not} an efficient a solution due to the off-chip memory's limited bandwidth.

Custom hardware implementations (e.g., ASICs) can potentially have as large on-chip memory as needed. For example, the GenASM ASIC~\citep{cali2020genasm} uses scratchpads of 96.5~$KiB$ each to hold the DP tables. However, these scratchpads occupy 76\% of the total chip area and consume over 54\% of the chip power. This limits the performance achievable with a given chip area and power budget.

In summary, GenASM has a large memory footprint compared to typical on-chip memory capacities in commodity hardware, and while sufficiently large on-chip scratchpads can be designed for custom hardware implementations, it is costly to do so.

\subsection{Scrooge} \label{sec:optimizations}
We have shown in \secref{challenges} that GenASM has a high amount of data movement \emph{and} high memory footprint per problem instance. We have elaborated that this combination either limits performance (on commodity hardware), or requires expensive large on-chip memories (on custom hardware), both of which are undesirable.
Thus, our strategy is to reduce the GenASM algorithm's memory footprint as much as possible while introducing minimal computational overhead. We present three novel algorithmic improvements that collectively achieve a 24\x\ reduction in memory footprint, as well as a 12\x\ reduction in data movement from the memory that holds the DP table.

\subsubsection{Improvement~1 - Store Entries, not Edges (SENE)} \label{sec:SENE}
\begin{figure}[h]
    \centering
    \includegraphics[width=\bitapfigwidth\columnwidth,keepaspectratio]{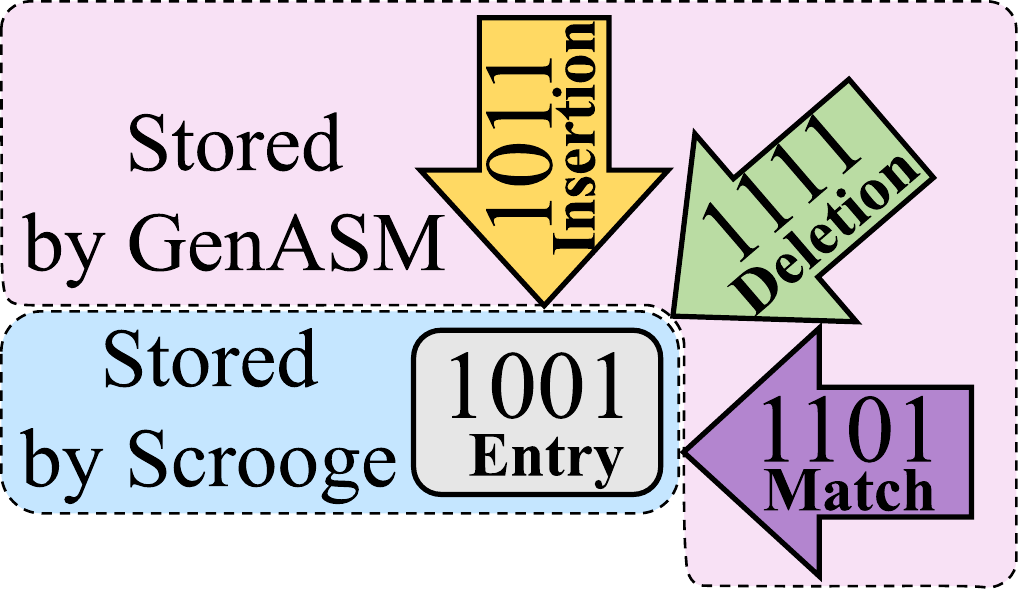}
    \caption{Per cell, GenASM stores three edges for traceback. \toolname with SENE stores only the DP entry itself instead; the needed edges are regenerated on the fly during traceback.}
    \label{fig:SENE}
\end{figure}

As we explain in \secref{traceback}, GenASM stores 3 bitvectors per entry of the table \var{R} to enable traceback. If we imagine a graph where the entries of \var{R} are nodes and the intermediate bitvectors are edges connecting their source and target entries, GenASM stores 3 ingoing edges for most nodes (\figref{SENE}).
We propose to trade off the majority of this memory footprint for a small increase in computation with the \defn{SENE} improvement. SENE regenerates the required edges on-demand during traceback from stored nodes (entries of table \var{R}) by applying the update rules in \algref{bitap} on requested neighbor entries.

\textbf{Cost and benefits.}
Since traceback explores only a single path across the table \var{R}, only $O(\var{W})$ edges are regenerated, making the overhead of this extra computation small compared to computing the table of $O(\var{W}^2)$ entries. Storing \var{R} requires storing $65\times65$ entries of 64 bits each, for a total of $33,800B\approx 33kiB$. The previous memory footprint was $96.5kiB$ as derived in \secref{memory_footprint}, yielding a $\frac{96.5}{33} = 2.92 \approx 3\times$ improvement in memory footprint. Since each of these locations is still only written to once during the construction of \var{R}, SENE also reduces the data movement from the memory that holds the DP table by 3\x.

\subsubsection{Improvement~2 - Discard Entries not used by Traceback~(DENT)} \label{sec:DENT}
The windowing heuristic (\secref{windowing}) mandates that traceback covers only the first $\var{W}-\var{O}$ characters of each window. This means that traceback never reads the table entries of the last $\var{O}$ characters in each window. %

\begin{figure}[h]
    \centering
    \includegraphics[width=\bitapfigwidth\columnwidth,keepaspectratio]{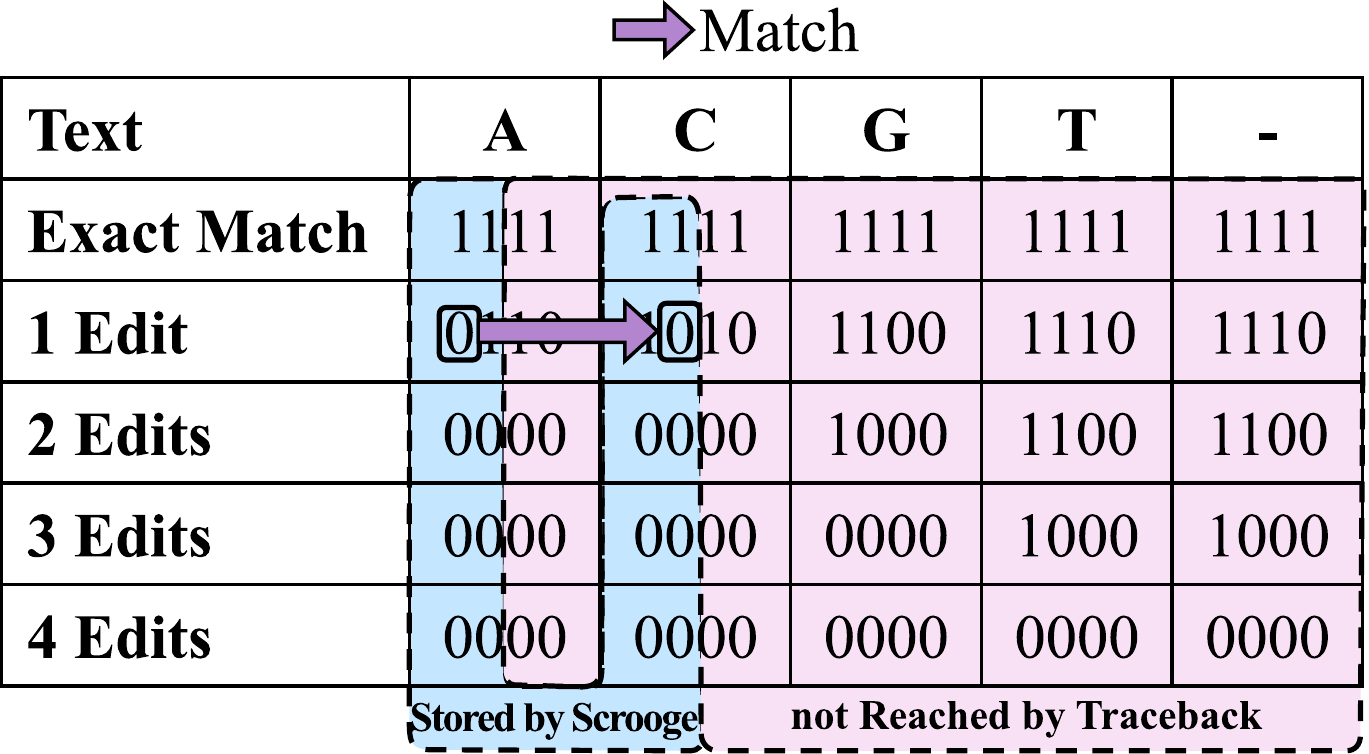}
    \caption{DENT exploits that the windowing heuristic stops traceback after the first $\var{W}-\var{O}$ edges are crossed (here $\var{W}$=4 and $\var{O}$=3). The area never reached by traceback can be discarded.}
    \label{fig:DENT}
\end{figure}

We propose to discard the entries that can never be reached by traceback, an improvement we call \defn{DENT}. These include the last $\var{O}$ columns of \var{R} and the last $\var{O-1}$ bits of every bitvector. The resulting DP table consists of $\var{W}-\var{O}+1$ columns, $\var{W}+1$ rows, and $\var{W}-\var{O}+1$ bits per entry. \figref{DENT} shows an example for $\var{W}$=4 and $\var{O}$=3, where \toolname stores only the leftmost 2 columns and leftmost 2 bits per entry, because traceback does not reach the rightmost 3 columns and rightmost 2 bits per entry.

Due to the fixed word sizes and word alignment requirements of commodity hardware, the number of bits stored for each bitvector cannot be chosen freely. We show in \secref{throughput} that for a modern GPU $\var{O}=33$ achieves the best throughput results for $\var{W}=64$, because the stored bitvectors perfectly fit into a 32-bit word. In contrast, \citep{cali2020genasm} uses \var{O}=24 for its ASIC design, which we show to be suboptimal on commodity hardware. Note that increasing \var{O} \emph{improves} accuracy, see \secref{windowing} for an intuition and \secref{accuracy} for experimental results.

\textbf{Cost and benefits.}
DENT incurs two computational overheads: First, the bits to store have to be determined and extracted from the bitvectors. Second, increasing \var{O} from 24 to 33 means the algorithm makes 9 characters less progress per window.

By discarding the right half of each bitvector and the rightmost \var{O} columns of \var{R}, DENT improves the memory footprint by $\frac{\var{W}}{\var{W} - \var{O}+1}\times\frac{\var{W}+1}{\var{W} - \var{O}+1} = \frac{64}{32}\times\frac{65}{32} \approx 4\times$. We describe in \S1 of the supplementary materials how DENT can be extended to store only half the rows of \var{R} for a total 8\x memory footprint reduction.

By the same calculation, the number of writes to table \var{R} is reduced by approximately 4\x, assuming the forefront diagonal (marked red in \figref{bitap_example}) is kept in registers and communicated directly.

\subsubsection{Improvement~3 - Early Termination (ET)} \label{sec:early_termination}
\begin{figure}[h]
    \centering
    \includegraphics[trim={1.2cm 0 0 0},width=\bitapfigwidth\columnwidth,keepaspectratio]{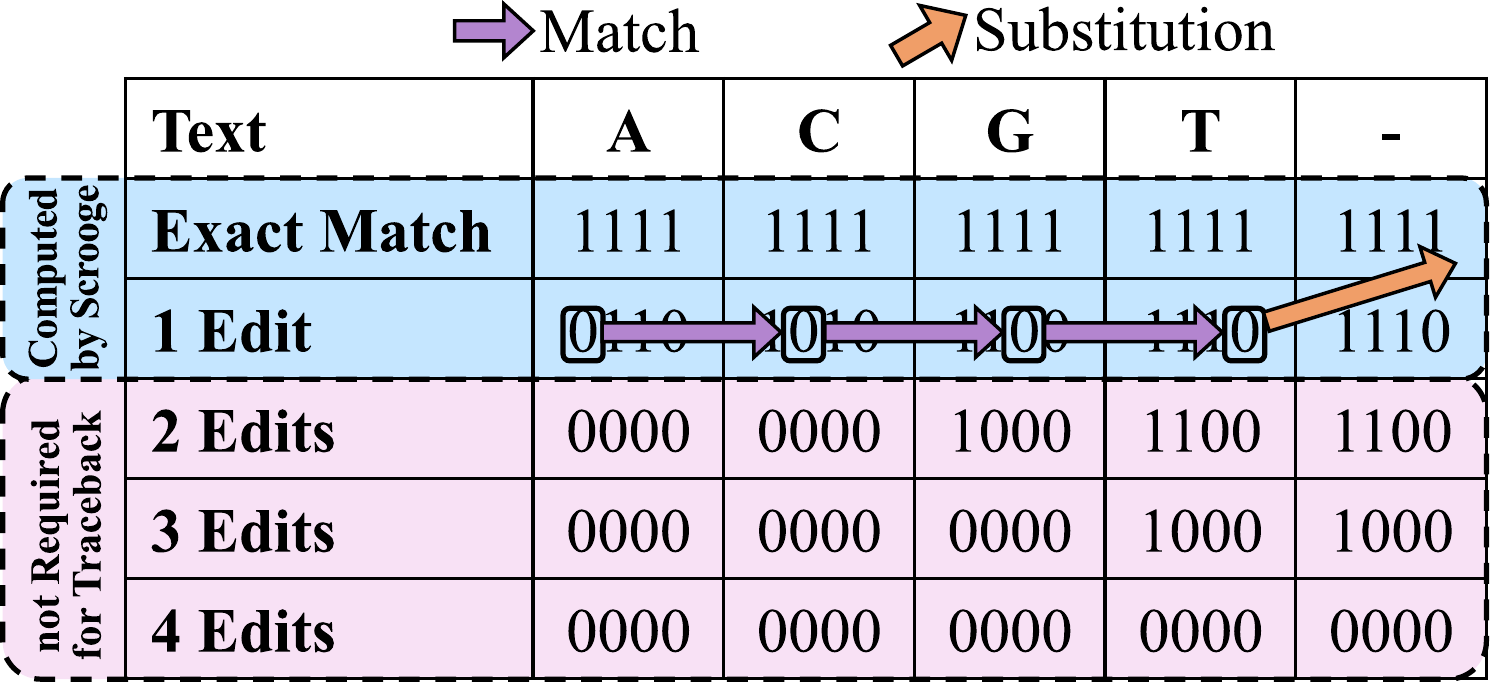}
    \caption{The colored edges indicate the path taken by traceback for \var{W}=4 and \var{O}=0. Rows below the edit distance do not contain useful information for traceback. Thus, they do not need to be computed (Early~Termination).}
    \label{fig:early_termination}
\end{figure}

The edit distance is determined by the highest row of \var{R} that contains a 0 in the most significant bit in the leftmost column. Traceback starts from this entry. Since entries are constructed from their north, north-east, and east neighbors, the traceback path can only go to the north, north-east, and east. It can \emph{never} go south. Thus, at no point do the rows of higher cost than $distance(\var{pattern}, \var{text})$ contain useful information for traceback (see \figref{early_termination} for an example). We propose building \var{R} row-wise, and terminating the algorithm early as soon as the most significant bit in the first entry of the current row is a 0.

\textbf{Cost and benefits.}
Early Termination does not yield a constant factor improvement in either memory footprint or runtime: If $distance(\var{pattern}, \var{text}) = \var{W}$, we are not able to terminate early at all. However, typical input pairs incur fewer than \var{W} edits in a single window. For correct candidate pairs, the edit distance will be low, e.g. up to 15\% for long reads~\citep{alser2020technology}. Even uncorrelated random sequence pairs of length \var{W} over a 4-letter alphabet have an edit distance of at most $\frac{3}{4}$\var{W} on average, as we prove in \S10 of the supplementary materials. Thus, on average \toolname can skip at least 25\% of the entries of \var{R}, saving computation as well as data movement (see \secref{data_movement_and_roofline_model}).

\textbf{Conflict with intra-task parallelism.}
Recall from \secref{task_parallelism} that GenASM provides the option for intra-task parallelism.
Exploiting this parallelism requires the available processing elements to build \var{R} in a \emph{diagonal-wise} fashion, as shown in \figref{bitap_example}. However, as we describe in \secref{early_termination}, to make full use of Early~Termination, \var{R} should be built \emph{row-wise}. As a compromise, we implement Early~Termination in a diagonal-wise fashion in our GPU implementation. As in the row-wise version, construction on \var{R} stops as soon as the leftmost processing element finds a 0 in the most significant bit. Due to the diagonal-wise computation, the other processing elements have already computed several rows ahead at this point, i.e. done unnecessary work. For this reason, the benefit of Early~Termination is limited in intra-task parallel implementations, such as our GPU implementation, while being much more significant in row-wise implementations, such as our CPU implementation. We reaffirm these effects experimentally in \secref{sensitivity}.

\subsection{Implementation} \label{sec:implementation}
We implement C++ versions of our algorithm for x86 CPUs and NVIDIA GPUs. They are exposed as simple library functions for pairwise sequence alignment. Each improvement and implementation constant can be easily configured at compile time through preprocessor macros. The implementations, as well as baselines and evaluation scripts, are available at \gitrepo.

\textbf{CPU.}
The CPU version converts the input pairs to a two-bit-per-basepair encoding, but padded to 8 bits. Each thread works on a single pairwise alignment at a time and obtains sequence pairs from a global queue. During each call to the GenASM-DC subroutine, the thread calculates the DP table \var{R} in a row-wise fashion.

\textbf{GPU.}
The GPU version is implemented using CUDA~11.1~\citep{cuda_guide} and targets GPUs of compute capability~7.0 and higher~\citep{cuda_gpus}. The input sequence pairs are converted to a two-bit-per-basepair encoding and transferred to the GPU. Each thread block works on a single pairwise alignment at a time and obtains sequence pairs from a global queue. During each call to the GenASM-DC subroutine, the thread block calculates the DP table \var{R} in a diagonal-wise fashion, and each of the \var{W} threads in the thread block calculates a single column of \var{R}. Threads resolve their mutual data dependencies using warp shuffle instructions within a warp and using shared memory across warps. A single thread per warp executes the traceback operation. The size of the CIGAR string is not known ahead of time, hence it is stored as a linked list in global memory.

\section{Results} \label{sec:results}
\subsection{Evaluation Methodology}
We demonstrate the benefits of Scrooge (along with each of our three algorithmic improvements) using both CPU and GPU implementations by comparing it to the recent  WFA lm~\citep{Eizenga2022wfalm}, WFA~\citep{wfa_algorithm}, KSW2~\citep{ksw2_a} (the state-of-the-art aligner used in minimap2~\citep{ksw2_b}), Edlib~\citep{edlib} (the state-of-the-art implementation of Myers' bitvector algorithm~\citep{myers_algorithm_1999} used in Medaka~\citep{medaka} and Dysgu \citep{cleal2022dysgu}), CUDASW++3.0~\citep{liu2013cudasw++}, Darwin-GPU~\citep{ahmed2020darwin_gpu}, and our CPU and GPU implementations of the GenASM algorithm.

We evaluate the throughput and accuracy of Scrooge via three classes of experiments. First, we compare the throughputs of all evaluated tools and show that Scrooge outperforms state-of-the-art aligners. Second, we evaluate the throughput benefits of \toolname's algorithmic improvements and its sensitivity to different choices for \var{W} and \var{O}. Third, we evaluate \toolname's accuracy. %
We define throughput as the number of pairwise sequence alignments per second for a given dataset.

We run all CPU evaluations on a dual-socket Intel Xeon Gold 5118 (2\x~12 physical cores, 2\x~24 logical cores)~\citep{xeon_gold_5118} at 3.2GHz with 196GiB DDR4 RAM. We run all GPU evaluations on an NVIDIA A6000~\citep{A6000}. We repeat all CPU and GPU experiments 10 times and 5 times, respectively, and average the results.

\subsubsection{Datasets}
We simulate 115,240 PacBio reads from the human genome using PBSIM2~\citep{pbsim2}, each of length 10 kilobases and with a target error rate of 5\%. We obtain the ground truth location in the reference genome, and the alignment (CIGAR string) of each read from PBSIM2, thus obtaining 115,240 candidate pairs for our \defn{long read groundtruth} dataset. We map 500 of the simulated PacBio reads to the human genome using minimap2~\citep{ksw2_b} and obtain all chains (candidate locations) it generates using the \texttt{-P} flag, 138,929 locations in total. This constitutes our \defn{long read} dataset. We map 100,000 Illumina short reads from the dataset with accession number SRR13278681 to the human genome using minimap2~\citep{ksw2_b} and obtain all chains (candidate locations) it generates using the \texttt{-P} flag, 9,612,222 locations in total. This constitutes our \defn{short read} dataset.
We show further statistics of the datasets in \S2 of the supplementary materials, including error, error rate, and sequence length distributions. The exact datasets and command lines that produced all our results, including those in the supplementary materials, are available at our GitHub repository: \gitrepo.

\subsection{Throughput}\label{sec:throughput}
We run the CPU-based tools using 48 threads. We set the band width (i.e., the edit distance threshold) of Edlib and KSW2 to 15\% of the read length. We configure WFA-adaptive as recommended by its authors. We take the fastest configuration from a parameter sweep for Darwin-GPU. For a meaningful comparison, we ensure that Darwin-GPU's alignment component fully aligns all sequence pairs. We explain our changes to Darwin-GPU in \S7 of the supplementary materials. We empirically configure \toolname's CPU and GPU implementations with \var{W}=64, \var{O}=33 for the long read dataset and \var{W}=32, \var{O}=17 for the short read dataset, and enable the combinations of improvements that yield the best throughput. The exact function calls and parameters we used for each tool can be found in our GitHub repository and in \S7 of the supplementary materials. \figref{performance_bars} shows that \toolname significantly speeds up the alignment of long and short reads over \emph{all} baselines. In particular, the CPU implementation of \toolname has \cpuoptspeedup higher throughput (i.e., pairwise sequence alignments per second) than our CPU implementation of GenASM for long reads and 3.8\x higher throughput for short reads. The GPU implementation of \toolname has \gpuoptspeedup higher throughput than our GPU implementation of GenASM for long reads and 2.4\x higher throughput for short reads. The CPU and GPU speedups over GenASM are entirely due to \toolname's algorithmic improvements (i.e., SENE, DENT, ET) since our \toolname and GenASM implementations are similarly optimized.

Note that WFA, KSW2, CUDASW++3.0, and Darwin solve a more general formulation of the alignment problem with affine gap scores~\citep{gotoh1982improved}. This puts them at a performance disadvantage. In contrast, Edlib~\citep{edlib}, GenASM~\citep{cali2020genasm}, and \toolname solve a less general but more efficient formulation of the alignment problem with unit costs (edit distance or Levenshtein distance~\citep{levenshtein1966binary}). We list the capabilities of each tool in \S6 of the supplementary materials.

\begin{figure}[h]
    \centering
    \includegraphics[width=\columnwidth,keepaspectratio]{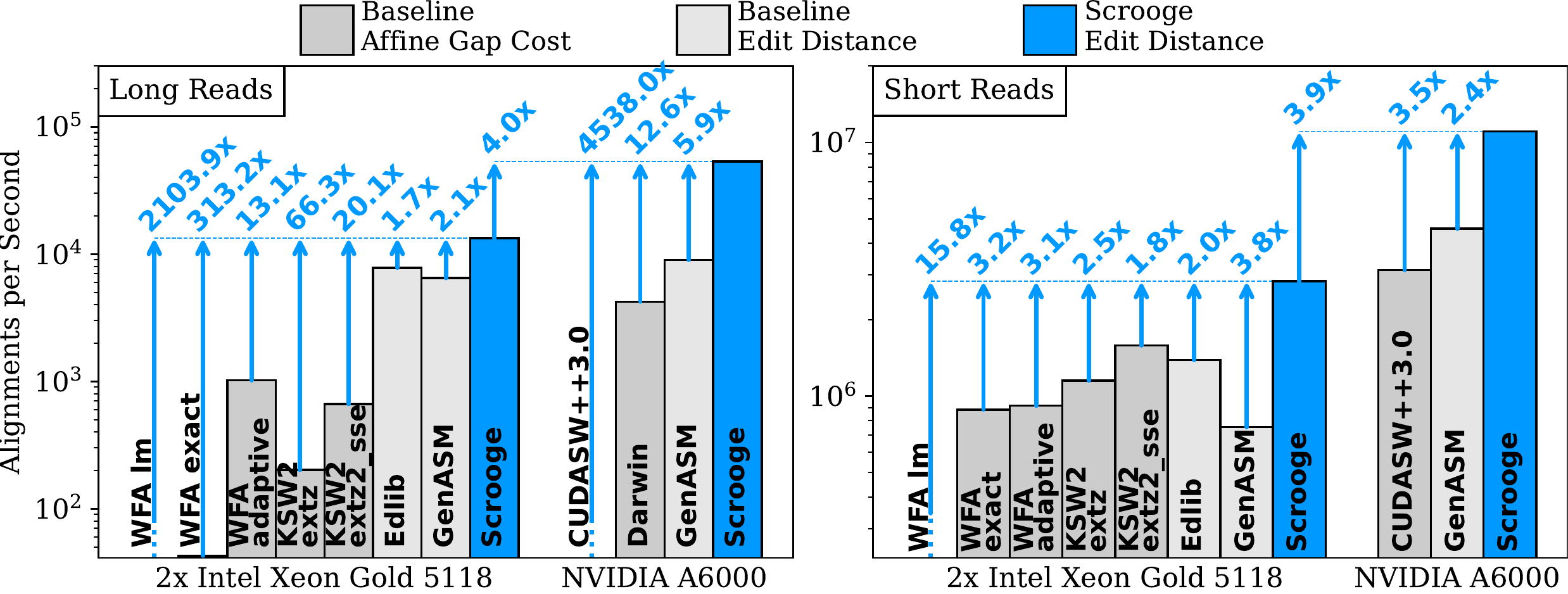}
    \caption{\toolname's alignment throughput relative to various CPU and GPU baselines.}
    \label{fig:performance_bars}
\end{figure}

\subsection{Thread Scaling}\label{sec:threadscaling}
\begin{figure}[h]
    \centering
    \includegraphics[width=\columnwidth,keepaspectratio]{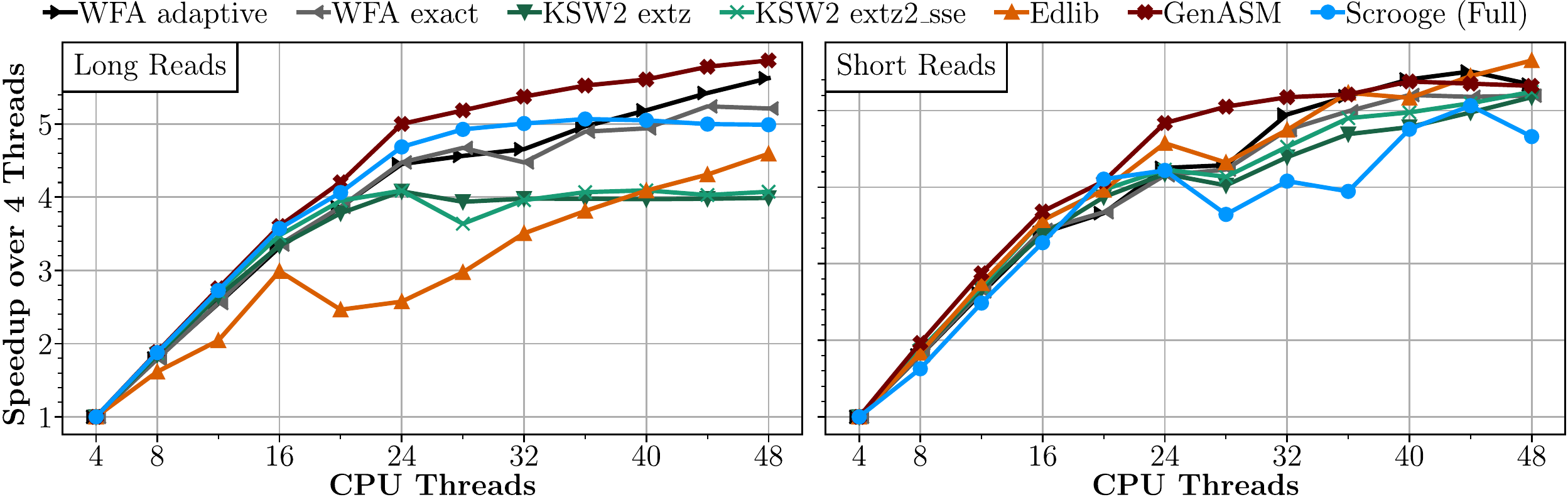}
    \caption{Speedup of each CPU tool as the number of CPU threads increases.}
    \label{fig:thread_scaling}
\end{figure}

We explore the scaling of each CPU tool as the number of CPU threads increases. For each evaluated CPU tool, we sweep the number of CPU threads and measure the throughput on the long read and short read datasets. \figref{thread_scaling} shows the results normalized to each tool's throughput with 4 threads (for readability). We make three key observations. First, most tools scale almost linearly up to 24 threads for both datasets, but do not scale significantly from 24 to 48 threads. The system we perform our experiments on has 24 physical cores and 48 logical cores~\citep{xeon_gold_5118}, thus we hypothesize that the tools do not benefit from simultaneous multithreading (\defn{Hyper-Threading} in Intel terminology)~\citep{hyperthreading} due to the low latencies of simple arithmetic and bitwise instructions~\citep{agner_instructiontables}, which is what the underlying alignment algorithms of the tools primarily consist of. Second, we observe that Edlib's performance \emph{decreases} from 16 to 20 threads in the long read dataset. Since this does not occur in the short read dataset, we hypothesize that Edlib suffers from cache thrashing in the long read dataset and that the data fits into the cache for the short read dataset. Third, we observe that both evaluated functions of KSW2 do not scale at all past 24 threads in the long read dataset. We hypothesize that KSW2 is bandwidth-bound in this case.

\subsection{Sensitivity Analysis}\label{sec:sensitivity}
We explore the throughput benefits of our algorithmic improvements in parameter sweeps over (1)~the number of GPU and CPU threads, (2)~the window size (\var{W}) parameter, and (3)~the window overlap (\var{O}) parameter.

\begin{figure}[h]
    \centering
    \includegraphics[width=\columnwidth,keepaspectratio]{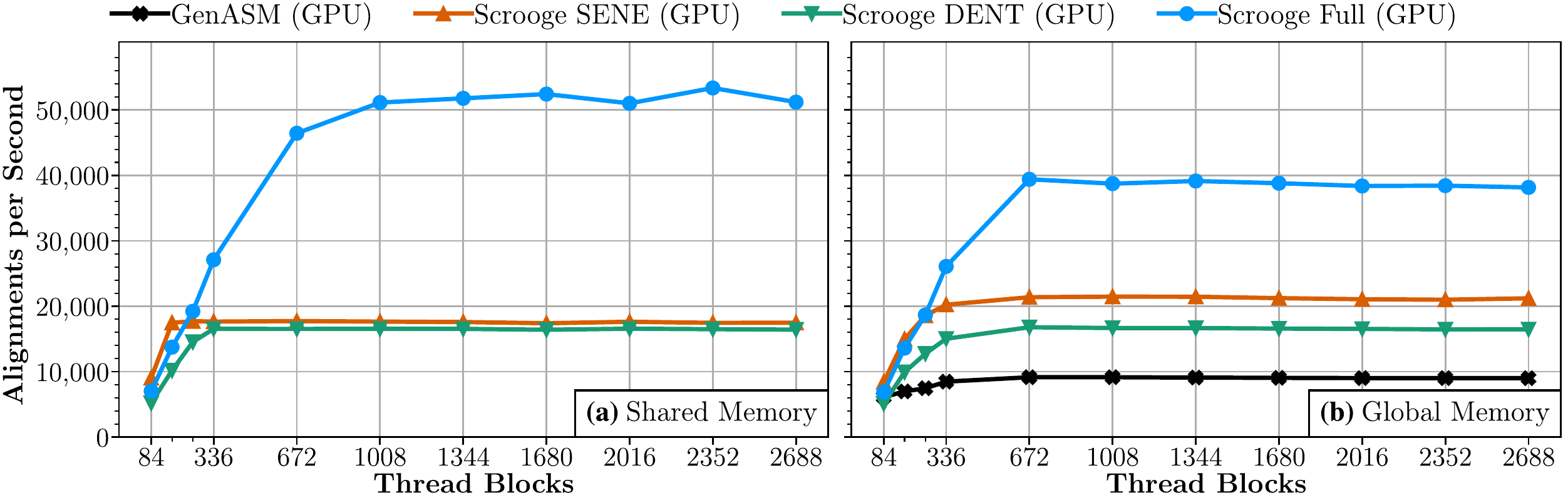}
    \caption{Scaling experiments of our GPU implementation with \var{W}=64, \var{O}=33, when the DP table placed in (a) shared memory and (b) global memory.}
    \label{fig:gpu_threadblocks}
\end{figure}

\textbf{GPU threads.} First, we run a scaling experiment on a GPU for GenASM, \toolname with the SENE improvement, \toolname with the DENT improvement, and \toolname with all three proposed improvements, with the DP table placed in either shared memory (\figref[a]{gpu_threadblocks}) or global memory (\figref[b]{gpu_threadblocks}).
Based on \figref{gpu_threadblocks}, we make five observations. First, we observe that SENE and DENT individually improve performance when the DP table is placed in either shared or global memory. Second, we observe that SENE, DENT, and Early Termination can be combined for greater benefits. Third, we observe that placing the DP table in shared  (on-chip) memory achieves the best performance, but only when both proposed memory footprint improvements (i.e., SENE and DENT) are applied. This is because only with SENE and DENT is the memory footprint small enough to keep sufficiently many problem instances in the shared memory to utilize the the compute resources in each SM (see \secref{memory_footprint}) well. Fourth, in configurations where the memory footprint is not reduced sufficiently (e.g., with only DENT or SENE), using global (off-chip) memory can be faster than using shared memory, because global memory has sufficient capacity to fit many problem instances, utilizing compute resources better than shared memory despite the global memory's limited bandwidth. Finally, we observe that the baseline GenASM algorithm cannot run using shared memory at all, although we showed in \secref{memory_footprint} that a single instance of the baseline DP table has a footprint of 98.5KiB and thus should use fit into the 99KiB of shared memory. This is because our implementation requires some additional memory, such as for communication between processing elements. Thus, we cannot fit even a single instance into shared memory with GenASM.

We ran the experiment for all seven possible combinations of our three improvements (i.e., SENE, DENT, and Early Termination). Full results are shown in \S11 of the the supplementary materials. In particular, we observe no significant benefits for GPUs from Early Termination, which is why we omit it in \figref{gpu_threadblocks} for readability.

\textbf{CPU threads.} We run a similar scaling experiment on a CPU for
GenASM, \toolname with the SENE improvement, \toolname with the Early Termination (ET)
improvement, and \toolname with the SENE and Early Termination improvements. From \figref[a]{cpu_threads_WO}, we make three observations: First, we observe that Early~Termination improves performance significantly. This contrasts with our GPU implementation, where Early Termination did not show significant benefits. This is because our CPU implementation builds the DP table \var{R} row-wise, while our GPU implementation builds \var{R} diagonal-wise (see \secref{early_termination}). Second, SENE improves performance consistently, but less significantly than in the GPU case. This is because modern CPUs have relatively large on-chip cache capacities (e.g., 1MiB L2 cache per core on the Xeon Gold 5118 we evaluated on~\citep{xeon_gold_5118}). Thus, the DP table easily fits into the L2 cache even without \toolname's algorithmic improvements, and hence reducing the memory footprint is not as important. Third, \toolname scales linearly up to 24 threads but does not scale at all from 24 to 48 threads, a trend we observe for all evaluated tools (see \secref{threadscaling}).

\begin{figure}[h]
    \centering
    \includegraphics[width=\columnwidth,keepaspectratio]{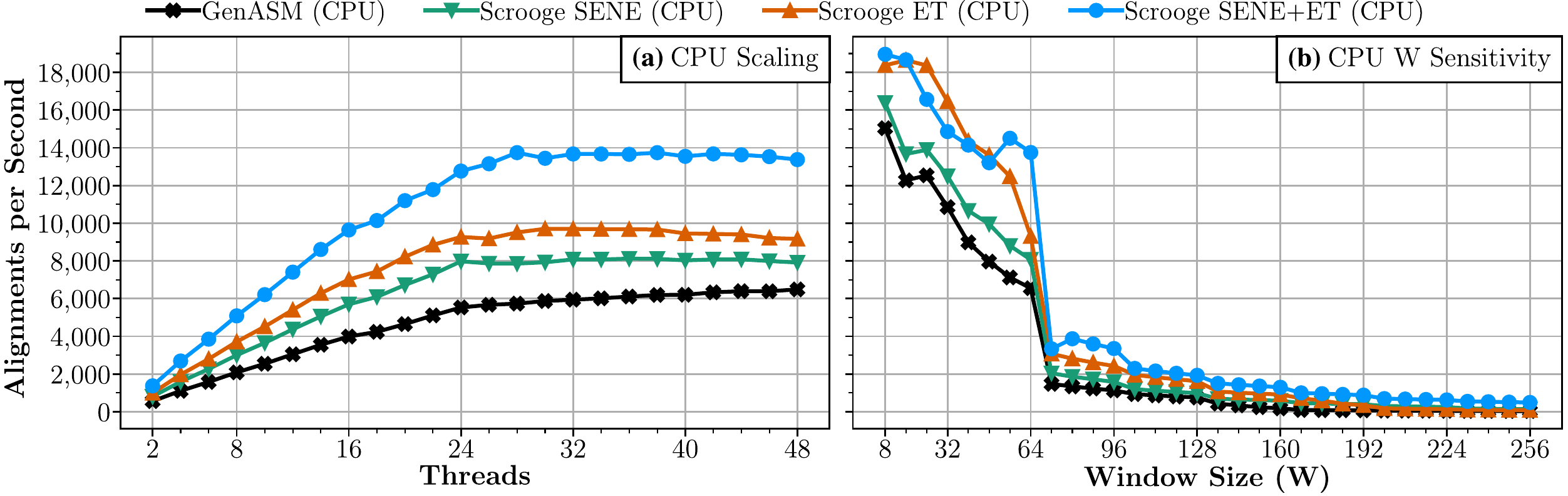}
    \caption{(a) Scaling and (b) sensitivity to window size of our CPU implementation. }
    \label{fig:cpu_threads_WO}
\end{figure}

We ran the experiment for all seven possible combinations of our three improvements (i.e., SENE, DENT, and Early Termination). Full results are shown in \S12 of the the supplementary materials. In particular, we observe no significant benefits for CPUs from DENT, and in some cases even a slowdown, which is why we omit it in \figref{cpu_threads_WO} for readability.

The three key takeaways from these experiments are that (1)~the SENE and DENT memory improvements yield significant benefits if performance is limited by memory bandwidth or capacity (e.g., in the GPU experiment), (2)~some of the algorithmic improvements can cause slight performance loss in practice (e.g., DENT in the CPU experiment), and (3)~the ideal combination of improvements depends on the computation platform (e.g., the available on-chip cache capacity) and the exact implementation (e.g., row-wise or diagonal-wise).

\textbf{Window size (\var{W}) and overlap (\var{O}).}
We explore the sensitivity of \toolname's throughput to the window size parameter \var{W} (\secref{windowing}) on CPUs. We vary \var{W} and set \var{O}=\var{W}//2+1. Note that larger \var{W} improves accuracy (\secref{windowing}).

From the CPU results in \figref[b]{cpu_threads_WO} we make two observations: First, we observe that performance generally reduces as \var{W} increases. This is because the number of calculated bits per window increases cubically with increasing \var{W}. Second, we observe a sudden throughput dropoff when \var{W} increases past 64. This is because the word size of the Xeon Gold 5118 CPU is 64 bits; thus, if \var{W}$>$64, each bitvector operation has to be emulated using multiple word-sized machine instructions. This emulation is conceptually simple (e.g., carry over shifted bits) but requires several additional instructions, causing the performance dropoff.
For example, in our implementation, a single 65-bit left shift is performed using two 64-bit left shifts, a 64-bit right shift, and a bitwise or operation.

We repeat the same study on a GPU and observe the same trends: Increasing \var{W} reduces performance, and if the bitvectors are longer than the machine word (32 bits on the evaluated GPU), bit operations become significantly more expensive. We plot the GPU results and give detailed explanations in \S4 of the supplementary materials.

We repeat a similar study for the window overlap (\var{O}) in \S5 of the supplementary materials. We observe that as \var{O} increases, performance generally reduces. However, with \toolname's optimizations, larger values of \var{O} can sometimes \emph{increase} performance. \var{O}=33 gives the best result. Thus, we choose it as the default operating point of \toolname\ for CPUs and GPUs.

\subsection{Area \& Power Consumption of an ASIC Implementation} \label{sec:hw_benefits}
The GenASM ASIC designed in \citep{cali2020genasm} uses a large on-chip scratchpad to store bitvectors for traceback. This scratchpad alone accounts for 0.256$mm^2$ (76\%) of silicon area and $0.055W$ (54\%) of power out of a total of 0.334$mm^2$ and $0.101W$ per accelerator core. Our proposed algorithmic improvements can be applied to that ASIC design through minor modifications. We estimate the area and power cost of such an ASIC implementation of \toolname analytically as follows:
\begin{enumerate}
    \item We start with the DC-logic, DC-SRAM, and TB-logic area and power numbers reported in \citep{cali2020genasm}
    \item We estimate \toolname's TB-SRAM area and power cost with CACTI~7~\citep{balasubramonian2017cacti}, as in \citep{cali2020genasm}, but with \toolname's reduced memory footprint and data movement numbers.
    \item We account for the logic overhead of SENE by adding the area and power of a single DC processing element~\citep{cali2020genasm} to the traceback (TB) logic cost, which accounts for recomputing edges during traceback. We assume no overhead for SENE during the construction of R, since the ANDed bitvectors are already computed.
    \item We assume no overheads for applying DENT since it simply masks out bits when storing the bitvectors, which is trivial in hardware.
\end{enumerate}

\tabref{asic_numbers} lists the area and power breakdowns obtained with this methodology, and the breakdown of \citep{cali2020genasm} as a comparison point. In particular, we observe a 3.6\x reduction in chip area and a 2.1\x reduction in chip power consumption, while maintaining the same throughput. These improvements come from (1)~the reduced TB SRAM capacity, and (2)~the reduced TB SRAM bandwidth. 

The key takeaway from this estimate is that \toolname's algorithmic improvements (1)~are directly applicable to and (2)~yield significant benefits over an ASIC implementation of GenASM.

\begin{table}[h]
\caption{Estimated area and power of a \toolname ASIC with \var{W=64} \var{O=33}.}
\label{table:asic_numbers}
\resizebox{\columnwidth}{!}{%
\setlength\tabcolsep{1.5pt}
\begin{tabular}{|l|rrrrr|rrrrr|}
    \hline
    & \multicolumn{5}{c|}{Area ($mm^2$)} & \multicolumn{5}{c|}{Power ($W$)} \\
    \textbf{ASIC Implementation} & \textbf{DC Logic} & \textbf{TB Logic} & \textbf{DC SRAM} & \textbf{TB SRAM} & \textbf{total} & \textbf{DC Logic} & \textbf{TB Logic} & \textbf{DC SRAM} & \textbf{TB SRAM} & \textbf{total} \\ \hline
    \citep{cali2020genasm} & 0.049 & 0.016 & 0.013 & 0.256 & 0.334 & 0.033 & 0.004 & 0.009 & 0.055 & 0.101 \\ \hline
    \toolname & 0.049 & 0.016 & 0.013 & 0.014 & 0.093 & 0.033 & 0.004 & 0.009 & 0.003 & 0.049 \\ \hline 
\end{tabular}}
\end{table}

\subsection{Accuracy} \label{sec:accuracy}
The GenASM algorithm~\citep{cali2020genasm}, which \toolname is based on, is a greedy heuristic algorithm, as explained in \secref{genasm_algo}. Our improvements do \emph{not} introduce additional inaccuracy. In fact, \toolname's default operating point of \var{W}=64 \var{O}=33 \emph{increases} accuracy (see \secref{windowing}) over GenASM's default operating point of \var{W}=64 \var{O}=24~\citep{cali2020genasm}. The following analysis explores the accuracy of both \toolname and GenASM across different operating points. At any given operating point, \toolname produces the same alignments (and hence accuracy) as GenASM at that operating point. We run three types of experiments. First, we evaluate the alignment quality of \toolname compared to all evaluated baseline tools. Second, we explore in detail the sensitivity of accuracy to the window size \var{W}. Third, we explore in detail the sensitivity of accuracy to the window overlap \var{O}.

\begin{figure}[h]
    \centering
    \includegraphics[width=\columnwidth,keepaspectratio]{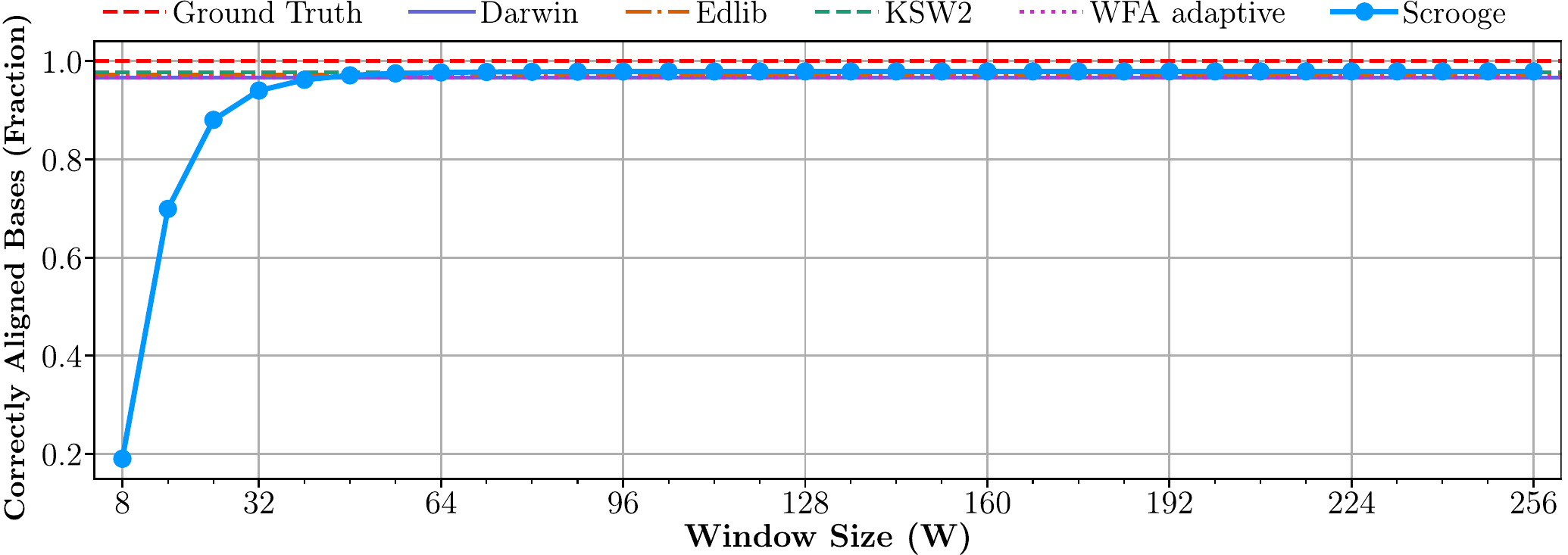}
    \caption{Fraction of correctly aligned bases according to the ground truth alignments in the long read groundtruth dataset.}
    \label{fig:matching_bases}
\end{figure}

\textbf{Alignment quality compared to baseline tools.}
We explore the quality of the alignments (CIGAR strings) generated by \toolname, compared to the baseline tools. To measure alignment quality, we count the number of correctly aligned bases according to the ground truth alignments reported by the PBSIM2 simulator for the long read groundtruth dataset. For \toolname we repeat the evaluation for multiple values of \var{W} and set \var{O}=\var{W}//2+1. We make three observations from \figref{matching_bases}. First, the number of bases correctly aligned by \toolname increases as the window size \var{W} increases. Second, \toolname correctly aligns approximately the same number of bases as all of the baselines if \var{W}$\geq$64. Third, no tool can consistently produce the exact ground truth alignment. By manually inspecting such misalignments of each tool, we determine this is because of two reasons. First, indels in homopolymers are ambiguous and cannot reliably be retrieved with any aligner. Second, sometimes the ground truth alignment is sub-optimal in terms of alignment score and/or edit distance. In these cases, the aligners' goal of finding the optimal scoring alignment produces high-scoring but wrong alignments.

We explore the sensitivity of \toolname's accuracy to the window size parameter \var{W} (\secref{windowing}). We analyze the accuracy compared to optimal edit distance solutions, such as Edlib~\citep{edlib}. We evaluate the generated alignments based on minimap2's default affine gap scoring model. We vary \var{W} and set \var{O}=\var{W}//2+1. For each experiment, we record the 0.5, 0.1, 0.01, and 0.001 percentile alignment scores (i.e., for a dataset of 1000 pairs, the 0.5 percentile would be the 500th worst alignment score, the 0.01 percentile would be the 10th worst alignment score) of \toolname\ and GenASM (which produce the same results for the same choice of \var{W} and \var{O}) and compare to Edlib as an ideal upper bound.

\begin{figure}[h]
    \centering
    \includegraphics[width=\columnwidth,keepaspectratio]{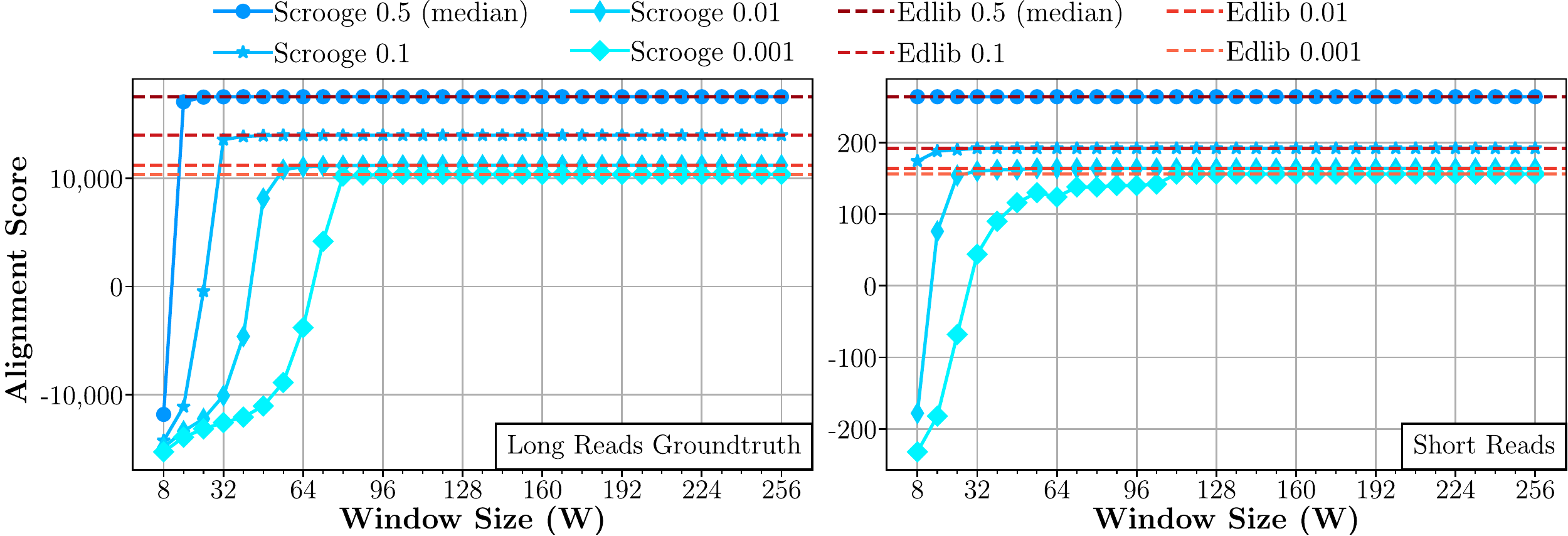}
    \caption{Sensitivity of \toolname's accuracy to \var{W}. We show the achieved alignment score of the 0.001, 0.01, 0.1, and 0.5 (median) quantiles, and compare to Edlib as an upper bound for the accuracy achievable with the edit distance metric.}
    \label{fig:accuracy_sweep_WO}
\end{figure}

\textbf{Sensitivity to window size (\var{W}).}
From \figref{accuracy_sweep_WO}, we make three observations: First, accuracy depends on the dataset. Second, small window sizes are sufficient for \toolname and GenASM to find the optimal edit distance alignment for \emph{most} of the sequence pairs. For example, the median alignment score is already optimal at \var{W}=32 for the long read groundtruth dataset and at \var{W}=8 for the short read dataset. Third, to find the optimal alignment for a few worst-case pairs, large window sizes are required: For example, the optimal alignment for the 0.001 percentile in the long read groundtruth dataset is only found for \var{W}$\geq$80. We manually inspect several of these "difficult" sequence pairs to find the reason for their apparent difficulty. We observe sequence pairs are aligned poorly if they contain extremely noisy and repetitive sub-sequences. However, these pairs \emph{will} be aligned optimally if the window size is larger than the length of the noisy sub-sequences. We illustrate this observation with an example sequence pair from the long read groundtruth dataset in \S13 of the supplementary materials.

\textbf{Sensitivity to window overlap (\var{O}).}
We explore the sensitivity of \toolname's accuracy to the window overlap parameter \var{O} (\secref{windowing}). We sweep over \var{O} and run experiments for each $\var{W}\in\{32,64,96,128\}$. For each experiment, we record the 0.01 percentile alignment score of \toolname/GenASM and compare to Edlib as an ideal upper bound.

\begin{figure}[h]
    \centering
    \includegraphics[width=\columnwidth,keepaspectratio]{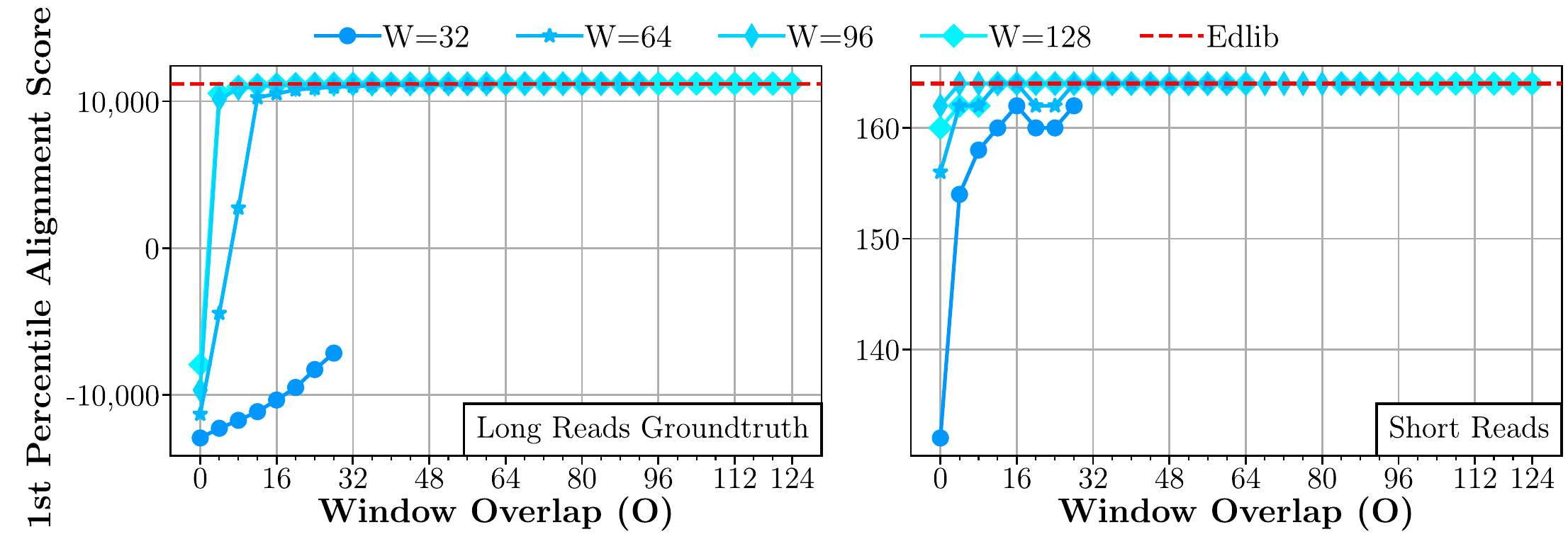}
    \caption{Sensitivity of accuracy to \var{O}, reporting the 1st percentile (worst 1\%) alignment score for each configuration. Edlib is an upper bound for the scores achievable with the edit distance metric.}
    \label{fig:accuracy_sweep_O}
\end{figure}

From \figref{accuracy_sweep_O}, we make two observations: First, accuracy improves as \var{O} increases. Second, we observe that \var{W} and \var{O} need to be \emph{balanced} to achieve good accuracy. For example, the accuracy loss of a too small \var{W}=32 for the long read groundtruth dataset cannot be overcome with even large \var{O}=30. Similarly, choosing \var{O} close to 0 hurts accuracy for both datasets, even when \var{W} is large.

The two key takeaways from these experiments are that (1)~\var{W} and \var{O} need to be chosen per dataset, and (2)~\var{W} and \var{O} should be increased or reduced together for the best accuracy.

\section{Discussion and Conclusion} \label{sec:discussion}
To our knowledge, this is the first paper to (1)~demonstrate the computational inefficiencies in the GenASM algorithm, (2)~address them with three improvements in our new \toolname algorithm, (3)~rigorously demonstrate the computational benefits of \toolname over GenASM for CPU, GPU, ASIC implementations, and (4)~rigorously analyze the accuracy of GenASM and \toolname under multiple different configurations.

We have already extensively compared to WFA~\citep{wfa_algorithm}, KSW2~\citep{ksw2_a, ksw2_b}, Edlib~\citep{edlib}, CUDASW++3.0~\citep{liu2013cudasw++}, and Darwin-GPU~\citep{ahmed2020darwin_gpu}. Several other works accelerate sequence alignment: NVBIO~\citep{nvbio} is a multi-purpose library for accelerating bioinformatics applications using GPUs, but is no longer maintained. Gasal2~\citep{gasal2} is a recent GPU aligner limited to short reads. CUDAlign4.0 \citep{de2016cudalign} can efficiently align a single pair of extremely long (chromosome-sized) sequences, with use cases such as whole genome alignment. Adept~\citep{awan2020adept} is a recent GPU aligner for short and long reads but does not support traceback, i.e. only reports the alignment score.

The Darwin accelerator~\citep{turakhia2019darwin} implements a Smith-Waterman-Gotoh accelerator for long reads using a similar greedy strategy to GenASM called \defn{tiling}. We have compared \toolname to the GPU implementation of this algorithm, Darwin-GPU. GenASM, \toolname, and Darwin demonstrate the significant benefits of greedy algorithms, based on which there are at least two interesting future directions to explore. First, a suitability study of different algorithms to greedy heuristics, such as Myers' bitvector algorithm~\citep{myers_algorithm_1999}, Hyyrö's banded bitvector algorithm~\citep{hyyro2003bit} or the recently proposed wavefront algorithm~\citep{wfa_algorithm}. Second, an exploration of the effectiveness of our algorithmic improvements for other implementations of greedy windowing or tiling, like Darwin. We believe the DENT improvement can be applied directly to Darwin.

We have demonstrated the computational benefits of \toolname over a variety of state-of-the-art baselines for both commodity hardware (i.e., CPUs and GPUs) and custom hardware (i.e., ASICs). We have demonstrated the accuracy of \toolname for multiple datasets. We conclude that \toolname\ has clear benefits across a wide range of computing platforms.

\section*{Funding}
We acknowledge the generous gifts of our industrial partners, especially Google, Huawei, Intel, Microsoft, VMware, Xilinx. This research was partially supported by the Semiconductor Research Corporation, the ETH Future Computing Laboratory, and the BioPIM project.

\renewcommand{\bibfont}{\small}
\balance
\bibliographystyle{IEEEtran}
\bibliography{main}

\cleardoublepage



\section*{Supplementary material}

# Supplementary Materials

## 1. DENT extended to rows

Since the W-O characters that are traced back can incur an edit cost of at most W-O, traceback also only reads at most W-O+1 rows starting from row distance(pattern, text). Thus the first max(0, distance(pattern, text) - (W-O+1)) do not need to be stored. Supplementary Figure 1 visualizes the concept of DENT extended to rows.

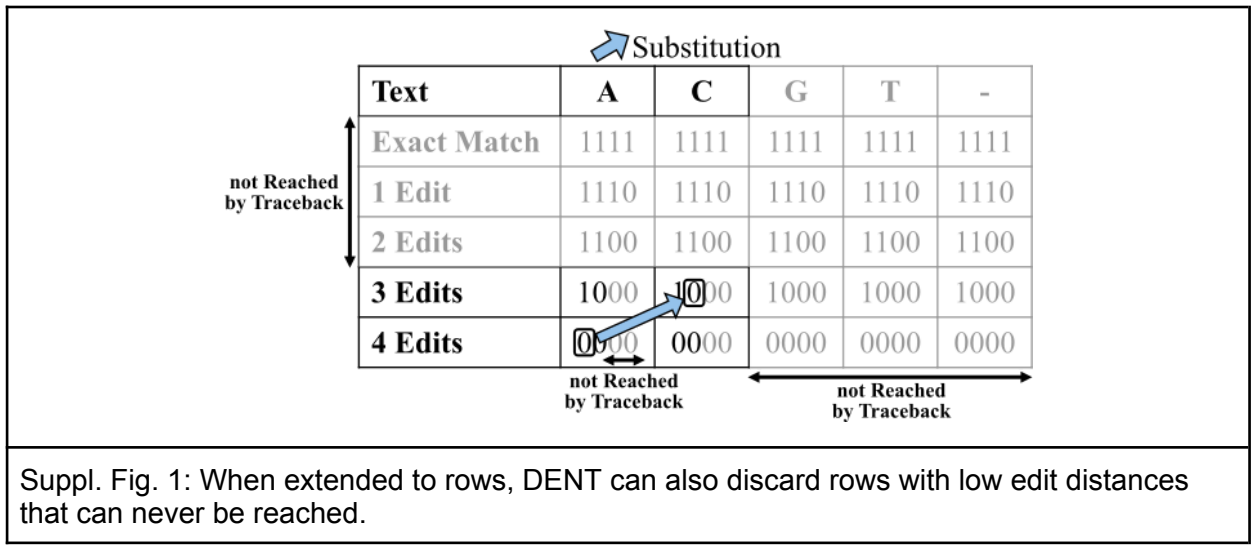

Suppl. Fig. 1: When extended to rows, DENT can also discard rows with low edit distances that can never be reached.

Note that determining the correct rows to store is dependent on distance(pattern, text), which is unknown ahead of the construction on R. If we build R row-wise we can always store the most recent W-O+1 rows of R and overwrite the oldest when we run out of space. To ensure we do not overwrite useful entries, we can apply the Early Termination improvement (Section 2.4.3) to stop as soon as we find distance(pattern, text).

Extended DENT (with SENE) provisions space for W-O+1 bits per entry, W-O+1 columns, and up to W-O+1 rows. For W=64 and O=33, this is $32^3$ bits. SENE without dent would provision $65^2*64$ bits, i.e., extended DENT improves the memory footprint by over 8x.

Extended DENT relies on row-wise computation and Early~Termination to not overwrite row entries required for traceback. This is not possible with the diagonal-wise computation, thus we only implement the weaker version of DENT, see Section 2.4.2. However, cases where extended DENT is useful can be easily imagined: Extended DENT needs row-wise computation, thus the program should be written in a sequential manner, as opposed to intra-task cooperative (which requires diagonal-wise computation). If that architecture then is constrained for on-chip memory capacity, extended DENT can address the issue.

## 2. Dataset Details

Supplementary Figures 2-7 show the edit and sequence length distributions for each dataset we used. The histograms were generated as follows:
1. Extract the sequence region pairs specified in the maf or paf file
2. Consider the length of the longer sequence in each pair as the pairs length
3. Align each pair with Edlib to obtain the edit distance, and divide it by the length of the pair
4. Draw the histograms

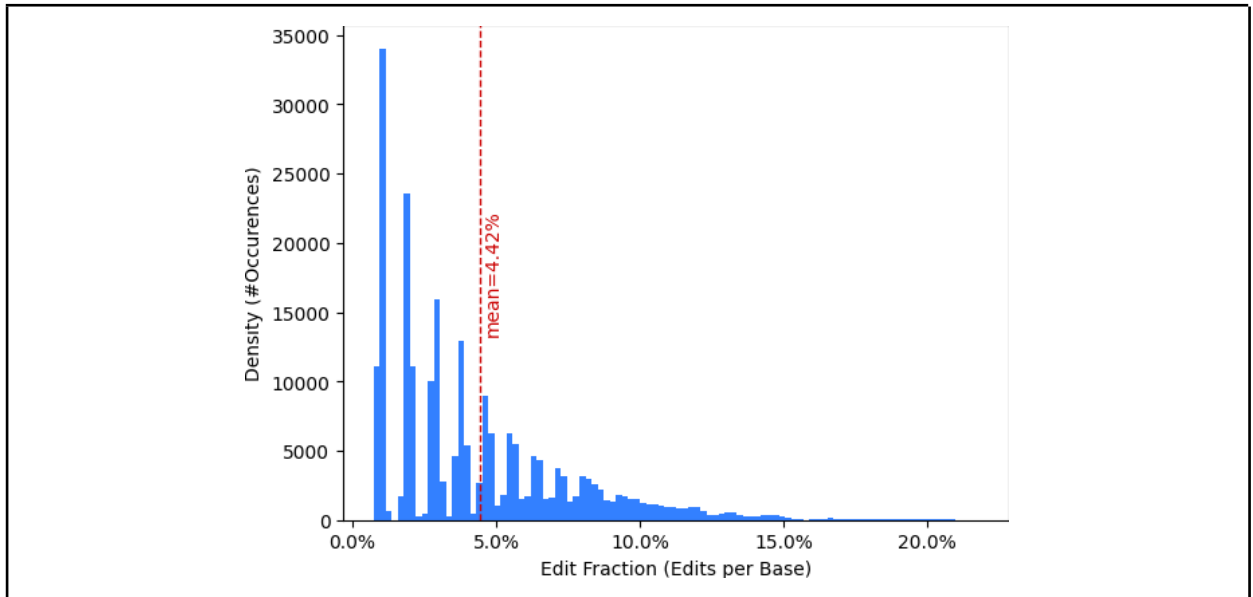

Suppl. Fig. 2: Histogram of edit distributions in the long read groundtruth dataset.

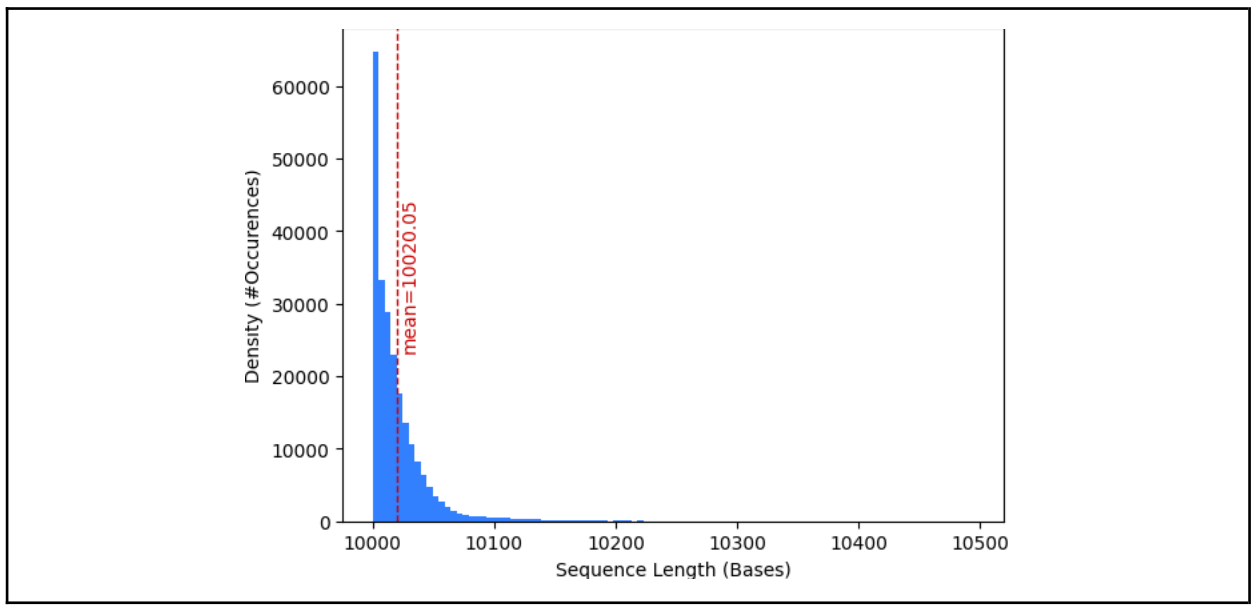

Suppl. Fig. 3: Histogram of the sequence lengths in the long read groundtruth dataset.

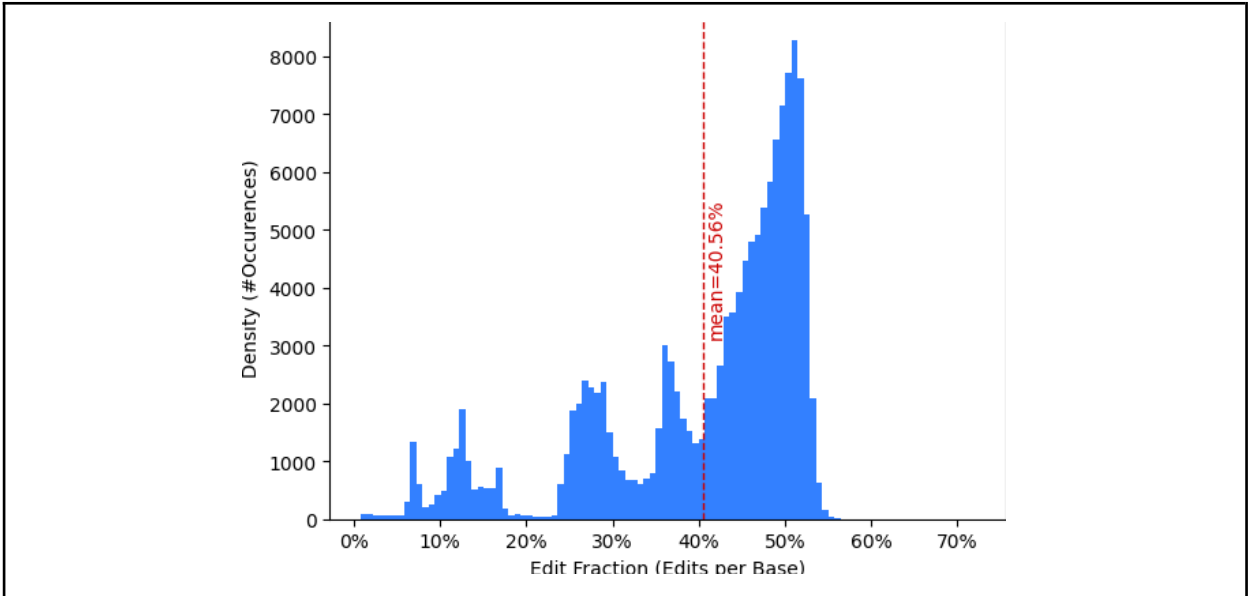

Suppl. Fig. 4: Histogram of edit distributions in the long read dataset.

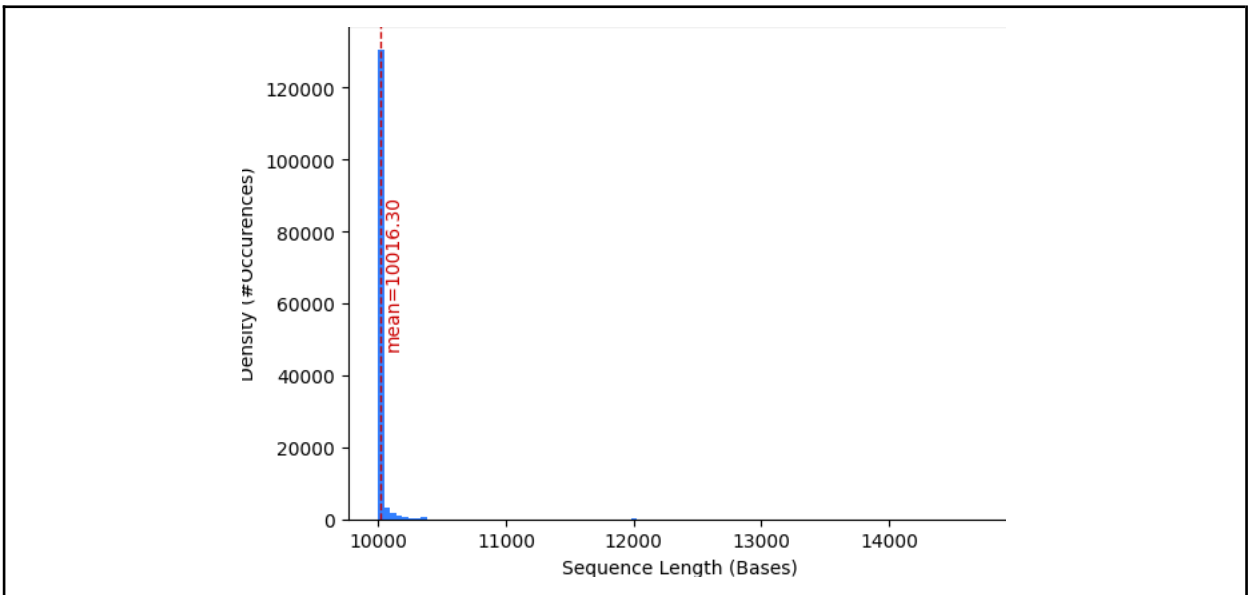

Suppl. Fig. 5: Histogram of the sequence lengths in the long read dataset.

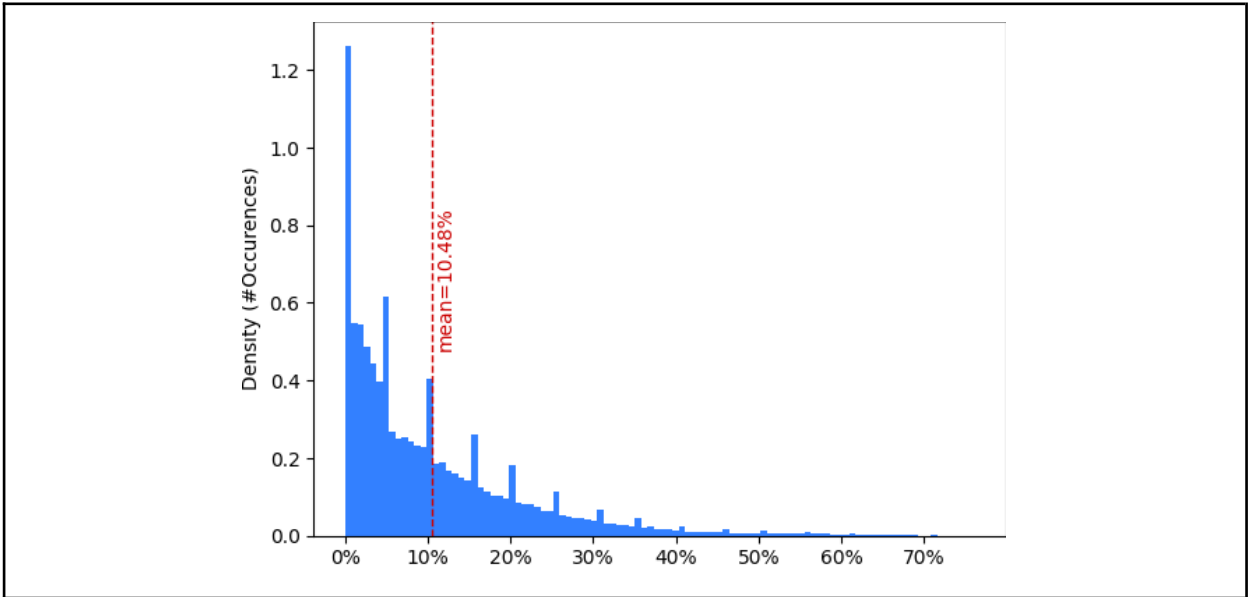

Suppl. Fig. 6: Histogram of edit distributions in the short read dataset.

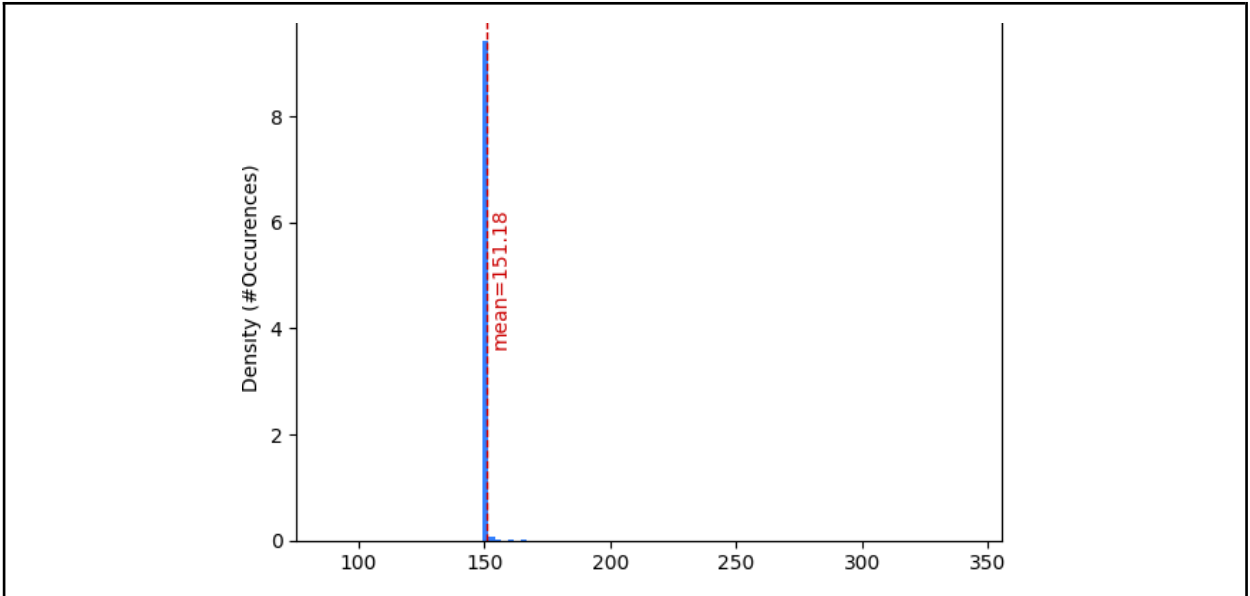

Suppl. Fig. 7: Histogram of the sequence lengths in the short read dataset.

# 3. Throughput Distributions

To understand the quality of our throughput results, we plot the throughput distributions of each evaluated tool as box plots. Supplementary Figure 8 shows the results. We observe that all throughput values are highly consistent.

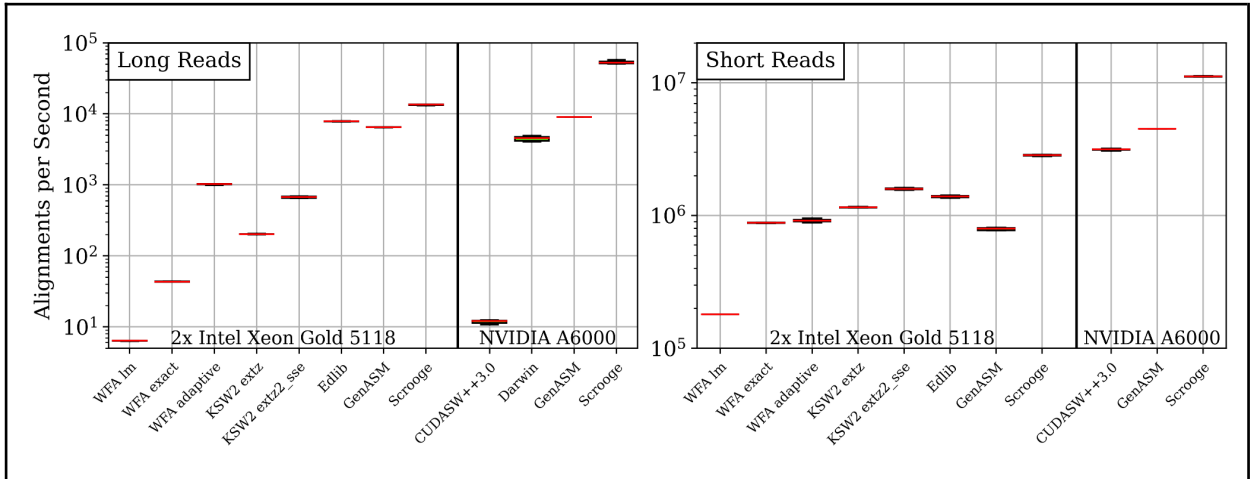

Suppl. Fig. 9: Throughput distributions of each evaluated tool. Each boxplot consists of the median as a red line, and the first and third quartiles as the lower and upper end of the box, respectively. The whiskers are 1.5 inter-quartile ranges down and up from the first and third quartile, respectively.

## 4. Throughput Sensitivity to Window Size (W) on GPU

We explore the sensitivity of Scrooge's throughput to the window size parameter W (Section 2.2.3) on GPUs. We vary W and set O=W//2+1. Note that larger W improve accuracy (Section 2.2.3). From the GPU results in Supplementary Figure 8 we make five observations: First, as on the CPU, performance generally reduces as W increases. Second, as on the CPU, there are sudden performance dropoffs whenever the window (and thus bitvector) size surpasses a multiple of the machine word size (32 bits on the tested GPU [4]). Third, some configurations can run only for small W when the DP table R is placed in shared memory. This is because as W increases, the memory footprint of R increases cubically, and some configurations run out of shared memory. With global memory, the DP tables can occupy as much of the GPUs off-chip memory as needed, thus Scrooge can run even very large W, although the performance eventually becomes impractically low. Fourth, for very small W, shared memory can offer significantly better performance than global memory. This is because for smaller W the memory footprint per DP table is smaller, and more instances can be kept in shared memory, thus easily achieving high occupancy. In contrast, global memory remains bound by memory bandwidth, because the operational intensity is unchanged. Fifth, for small window sizes and shared memory it is beneficial to run without the memory improvements SENE and DENT. This corroborates the observation made for the CPU implementation: If memory capacity and bandwidth are not limiting factors, then the computational overheads of SENE and DENT overshadow their benefits. In contrast, when W is large, the memory improvements become increasingly important because of the inverse argument. The three key takeaways from these experiments are that (1) the ideal combination of improvements depends on the window size (W) parameter, (2) the memory improvements SENE and DENT memory improvements are increasingly beneficial as the window size increases, and (3) the window size W should be exactly or a small multiple of the machine word size for optimal throughput.

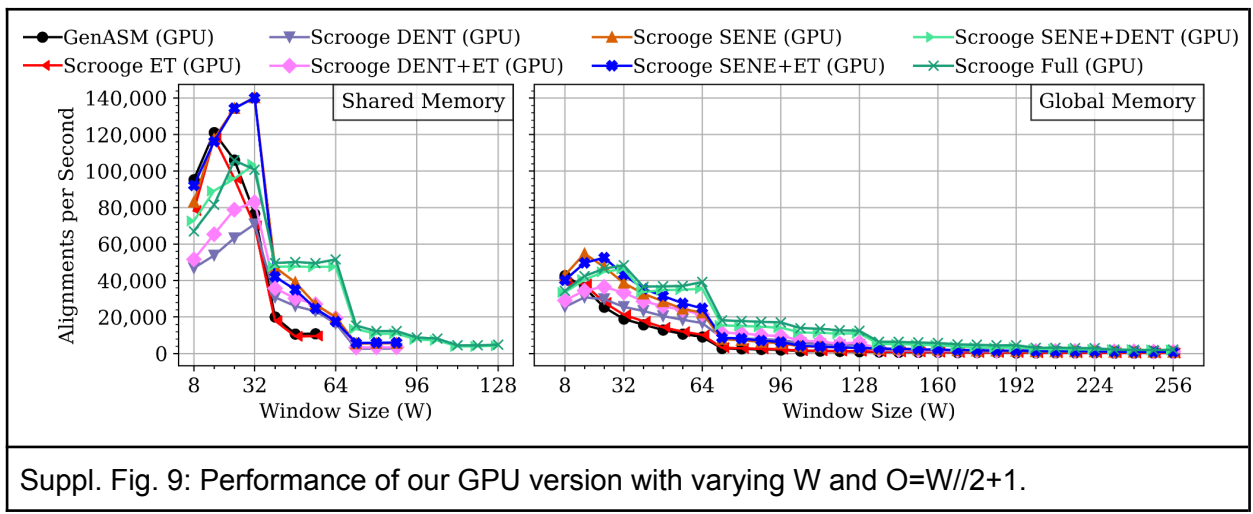

Suppl. Fig. 9: Performance of our GPU version with varying W and O=W//2+1.

## 5. Throughput Sensitivity to Window Overlap (O) on GPU

We explore the sensitivity of Scrooge's throughput to the window overlap parameter O (Section 2.2.3) on GPUs. We vary O and set W=64. Note that larger O improve accuracy (Section 2.2.3). From the GPU results in Supplementary Figure 9 we make two observations: First, when DENT is disabled, performance decreases strictly and smoothly as O increases. This is because the algorithm makes W-O characters progress per window, i.e. progress per window decreases linearly with increasing O. Second, we observe that with DENT enabled, performance can sometimes
increase as O increases, up to some limit. This is because DENT reduces the memory footprint of the DP table more as O increases. We observe a large and sudden increase in performance at O=34. This is because with DENT, each bitvector stored for traceback has size W-O+1, thus for O≥33 each stored bitvector fits into a single 32-bit machine word [3]. The key takeaway from this experiment is that a larger window overlap (O) can improve performance up to some limit. This is convenient because larger values of O also improve accuracy. In particular, we found for W=64 the optimal performance is attained by O=33, increasing both throughput and accuracy over the default operating point chosen by [6] of W=64 and O=24 for their accelerator.

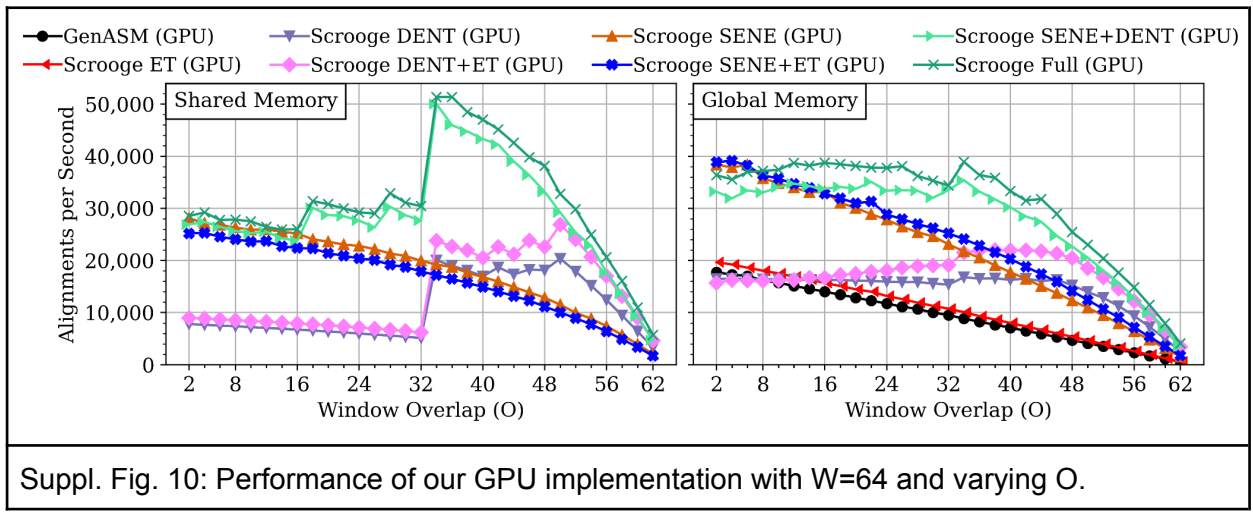

Suppl. Fig. 10: Performance of our GPU implementation with W=64 and varying O.

# 6. Tool Capabilities

Supplementary Table 1 lists the capabilities (scoring scheme, optimality guarantee, acceleration) of each evaluated tool variant.

| Suppl. Tab. 1: Properties of each evaluated tool variant. | | | | |
|---|---|---|---|---|
| **Tool** | **Supported Scoring Scheme** | **Banded?** | **Optimality Guarantee?** | **Acceleration** |
| WFA lm | One-piece Affine Gap | No | Yes | |
| WFA exact | One-piece Affine Gap | No | Yes | SSE2 |
| WFA adaptive | One-piece Affine Gap | No | No | SSE2 |
| KSW2 extz | One-piece Affine Gap | Yes | Yes | |
| KSW2 extz2_sse | One-piece Affine Gap | Yes | Yes | SSE2 |
| Edlib | Edit Distance | Yes | Yes | |
| GenASM (CPU) | Edit Distance | No | No | |
| Scrooge (CPU) | Edit Distance | No | No | |
| CUADSW++3.0 | One-piece Affine Gap | No | Yes | GPU |
| Darwin (GPU) | One-piece Affine Gap | No | No | GPU |
| GenASM (GPU) | Edit Distance | No | No | GPU |
| Scrooge (GPU) | Edit Distance | No | No | GPU |

# 7. Tool Function Calls and Parameters

Supplementary Table 2 lists the exact function calls and parameters we used to evaluate each tool.

| Tool | Clear Text Parameters | Function Call |
|---|---|---|
| \multicolumn{3}{l}{Suppl. Tab. 2: Function calls and parameters for each evaluated tool variant.} |
| WFA lm | Scores=1,1,1,1 | ```cpp
wfalm::SWGScores scores;
scores.match = 1;
scores.gap_extend = 1;
scores.gap_open = 1;
scores.mismatch = 1;
wfalm::wavefront_align_low_mem(text, pattern, scores);
``` |
| WFA exact | Scores=0,1,1,1 | ```cpp
affine_penalties_t affine_penalties = {
    .match = 0,
    .mismatch = 1,
    .gap_opening = 1,
    .gap_extension = 1
};
mm_allocator_t* mm_allocator = mm_allocator_new(BUFFER_SIZE_8M);
affine_wavefronts_t* affine_wavefronts
    affine_wavefronts_new_complete(max_read_length,
    max_read_length*105/100, &affine_penalties, NULL, mm_allocator);
affine_wavefronts_align(affine_wavefronts, inputs[pair_idx].pattern,
    inputs[pair_idx].m, inputs[pair_idx].text, inputs[pair_idx].n);
``` |
| WFA adaptive | Scores=0,1,1,1<br>Min_Wavefront_Length=10<br>Max_Distance_Threshold=50 | Runtime Evaluation:<br>```cpp
affine_penalties_t affine_penalties = {
    .match = 0,
    .mismatch = 1,
    .gap_opening = 1,
    .gap_extension = 1
};
const int min_wavefront_length = 10;
const int max_distance_threshold = 50;
mm_allocator_t* mm_allocator = mm_allocator_new(BUFFER_SIZE_8M);
affine_wavefronts_t* affine_wavefronts =
    affine_wavefronts_new_reduced(max_read_length,
    max_read_length*105/100, &affine_penalties, min_wavefront_length,
    max_distance_threshold, NULL, mm_allocator);
affine_wavefronts_align(affine_wavefronts, inputs[pair_idx].pattern,
    inputs[pair_idx].m, inputs[pair_idx].text, inputs[pair_idx].n);
``` |

|  | Scores=0,4,4,2<br>Min_Wavefront_Length=10<br>Max_Distance_Threshold=50 | Accuracy Evaluation:<br>```c
affine_penalties_t affine_penalties = {
    .match = 0,
    .mismatch = 4,
    .gap_opening = 4,
    .gap_extension = 2
};
const int min_wavefront_length = 10;
const int max_distance_threshold = 50;
mm_allocator_t* mm_allocator = mm_allocator_new(BUFFER_SIZE_8M);
affine_wavefronts_t* affine_wavefronts =
    affine_wavefronts_new_reduced(max_read_length,
    max_read_length*105/100, &affine_penalties, min_wavefront_length,
    max_distance_threshold, NULL, mm_allocator);
affine_wavefronts_align(affine_wavefronts, inputs[pair_idx].pattern,
    inputs[pair_idx].m, inputs[pair_idx].text, inputs[pair_idx].n);
``` |
| KSW2 extz | Scores=2,4,4,2<br>Band Width=15% | ```c
int8_t a = 2;
int8_t b = -4;
int8_t score_matrix[25] = { a,b,b,b,0, b,a,b,b,0, b,b,a,b,0, b,b,b,a,0,
    0,0,0,0,0 };
int gap_open = 4;
int gap_extend = 2;
ksw_extz(NULL, inputs[pair_idx].m, inputs[pair_idx].pattern,
    inputs[pair_idx].n, inputs[pair_idx].text, 5, score_matrix,
    gap_open, gap_extend, (int)(0.15*inputs[pair_idx].m), -1,
    KSW_EZ_EXTZ_ONLY, &ez);
``` |
| KSW2 extz2_sse | Scores=2,4,4,2<br>Band Width=15% | ```c
int8_t a = 2;
int8_t b = -4;
int8_t score_matrix[25] = { a,b,b,b,0, b,a,b,b,0, b,b,a,b,0, b,b,b,a,0,
    0,0,0,0,0 };
int gap_open = 4;
int gap_extend = 2;
int end_bonus = 0;
ksw_extz2_sse(NULL, inputs[pair_idx].m, inputs[pair_idx].pattern,
    inputs[pair_idx].n, inputs[pair_idx].text, 5, score_matrix,
    gap_open, gap_extend, (int)(0.15*inputs[pair_idx].m), -1,
    end_bonus, KSW_EZ_EXTZ_ONLY, &ez);
``` |
| Edlib | Band Width=15% | ```c
EdlibAlignConfig conf;
conf.k = (int)(0.15*inputs[pair_idx].m);
conf.mode = EDLIB_MODE_SHW;
conf.task = EDLIB_TASK_PATH;
conf.additionalEqualities = NULL;
conf.additionalEqualitiesLength = 0;
EdlibAlignResult res = edlibAlign(inputs[pair_idx].pattern,
    inputs[pair_idx].m, inputs[pair_idx].text, inputs[pair_idx].n,
    conf);
``` |

| | | |
|---|---|---|
| GenASM (CPU) | W=64 O=33 | ```
#define W 64
#define O 33
genasm_cpu::align_all(reference, reads, threads, &core_algorithm_ns);
``` |
| Scrooge (CPU) | W=64 O=33<br>SENE=True<br>DENT=False<br>EarlyTerm.=True | ```
#define W 64
#define O 33
#define STORE_ENTRIES_NOT_EDGES
#define EARLY_TERMINATION
genasm_cpu::align_all(reference, reads, threads, &core_algorithm_ns);
``` |
| CUADSW++3.0 | Defaults | Standalone |
| Darwin (GPU) | T=320 O=120 (Default) | Standalone<br>By default, Darwin-GPU does not fully evaluate or align sequence pairs whose score drops to 0. While sensible in the context of a full mapper, for a fair comparison of pairwise sequence alignment throughput, we modified Darwin-GPU's pairwise sequence alignment function to fully evaluate each sequence pair. The relevant change can be found at https://github.com/CMU-SAFARI/Scrooge/blob/main/baseline_algorithms/darwin-gpu/cuda_header.h#L177 |
| GenASM (GPU) | W=64 O=33 | ```
#define W 64
#define O 33
genasm_cpu::align_all(reference, reads, &core_algorithm_ns);
``` |
| Scrooge (GPU) | W=64 O=33<br>SENE=True<br>DENT=True<br>EarlyTerm.=False | ```
#define W 64
#define O 33
#define STORE_ENTRIES_NOT_EDGES
#define DISCARD_ENTRIES_NOT_USED_BY_TRACEBACK
genasm_cpu::align_all(reference, reads, &core_algorithm_ns);
``` |

# 8. Roofline Model Calculations

We analyze a single GenASM window of size W=64 (see Section 2.2.3). Since GenASM simply repeats the calculation of a window in a loop, this analysis is representative of the entire program.

## 8.1 GenASM Data Movement

Per window, GenASM generates W=64 intermediate bitvectors to initialize the topmost row of R (Line 11 of Algorithm 1). It also generates 4×64×64 intermediate bitvectors for the inner entries of R (Lines 13-17 of Algorithm 1), out of which 3×64×64 are stored for traceback (Section 2.2.2). Bitvectors have size W=64 bits each. We assume that the processing elements communicate neighbor entry values without memory accesses. This is a realistic assumption, for example, both the hardware accelerator in [6], as well as our GPU implementation are implemented this way. Thus, each of these bitvectors is written to memory exactly once per window (during the construction of R) and not read again. We ignore the memory accesses for reading the input sequences and for traceback, as the consumed bandwidth is negligible relative to DP table construction. This yields data_movement = (64 + 3×64×64) × 64b = 98,816B = 96.5KiB.

## 8.2 GenASM Operational Intensity

The operational intensity is defined as the ratio of work done over data movement [5]. We define work as the number of arithmetic and logic 64-bit integer operations (op), and data_movement as the number of bytes moved between registers and the memory hierarchy and scratchpad memory.
In each window, first, GenASM computes 65 entries to initialize the rightmost column, using 1 op each (Line 5). Second, it computes 64 entries to initialize the topmost row, using 2 op each (Line 11). Third, it computes 64×64 entries using 7 op each (Lines 13-17 of Algorithm 1). Thus, work = 65op + 64×2op + 64×64×7op = 28,865op. The GenASM algorithm's operational intensity is then I = work / data_movement ≈ 0.29 op/B .

## 8.3 Bandwidth and Compute Rooflines

**NIVIDIA A6000 GPU.** We abbreviate "streaming multiprocessor" as "SM".
We obtain the A6000's *global* (off-chip) memory bandwidth from its datasheet, 768 GB/s.
We derive the *shared* (on-chip) memory bandwidth from the CUDA Programming Guide [4] by multiplying the per-cycle-per-SM throughput (128B/cycle/SM)
with the clock frequency (1.8GHz) and number of SMs (84), 19353.6GB/s.
We derive compute throughput from the CUDA Programming Guide [4] by multiplying the per-cycle-per-SM throughput (32op/cycle/SM) with the clock frequency (1.8GHz) and number of SMs (84), 4838.4 Gop/s.

**Intel Xeon 5118 CPU.** We obtain the off-chip (DRAM) bandwidth by assuming 2,400MT/s DDR4 memory, multiplied by the bus width (8B) and maximum number of memory channels supported by the CPU (6), 115.2GB/s.

We obtain the cache bandwidths from the Intel Optimization Manual ([2]) by multiplying the respective per-cycle-per-core throughputs (L3: 15B/cycle/core, L2: 52B/cycle/core, L1: 133B/cycle/core) with frequency (2.3GHz) and number of cores (12), L3: 414GB/s, L2: 1,435.2GB/s, L1: 3,670.8GB/s.

We obtain the compute throughputs from the Intel Optimization Manual ([2]) by multiplying the respective per-cycle-per-core throughputs (Scalar: 4op/cycle/core, AVX512: 16op/cycle/core) with frequency (2.3GHz) and number of cores (12), Scalar: 110.4Gop/s, AVX512: 441.6Gop/s.

## 9. GenASM Memory Footprint Calculation

GenASM stores W=64 intermediate bitvectors to initialize the topmost row of R (Line 11 of Algorithm 1). It also generates 4×64×64 intermediate bitvectors for the inner entries of R (Lines 13-17 of Algorithm 1), out of which 3×64×64 are stored for traceback (Section 2.2.2). Bitvectors have size W=64 bits each. This yields working_set_memory_footprint = (64 + 3×64×64) × 64b = 98,816B = 96.5KiB.

## 10. Expected Edit Distance for Random String Pairs

**Lemma S1.** The expected edit distance between an uncorrelated random string pair of length W over an alphabet of size $|\Sigma|$ is at most $W*(|\Sigma|-1)/|\Sigma|$.
**Proof.** We prove Lemma S1 in two steps. First, we calculate the expected Hamming distance between an uncorrelated random string pair. Second, we prove that the edit distance of a string pair of equal length is at most its Hamming distance.

From Lemma S1 follows in particular that for a 4-character DNA alphabet, the expected edit distance between an uncorrelated random string pair of length W is at most W*(4-1)/4=3*W/4.

**Lemma S2.** The expected Hamming distance between an uncorrelated random string pair of length W over an alphabet of size $|\Sigma|$ is $W*(|\Sigma|-1)/|\Sigma|$.
**Proof.** The Hamming distance between two strings is defined as the number of locations where they differ [1]. For two uncorrelated random strings over an alphabet of size $|\Sigma|$, the probability that they differ at a given location is $P_{differ}=(|\Sigma|-1)/|\Sigma|$, and the expected number of differences at a single location is $E[differ]=P_{differ}=(|\Sigma|-1)/|\Sigma|$. For W possible locations in two strings of length W is then $E[differeces]=W*E[differ]=W*(|\Sigma|-1)/|\Sigma|=E[Hamming]$.

**Lemma S3.** The Hamming distance of a pair of strings of equal length is an upper bound on its edit distance.
**Proof.** The edit distance of a string pair is defined as the smallest number of single-character insertions, deletions, and substitutions to convert one into the other [7]. Suppose the Hamming distance of an equal-length string pair was strictly smaller than its edit distance. Then the edit distance could not have been the smallest number of single-character insertions, deletions, and substitutions to convert one string into the other, as the series of substitutions that resulted in the hamming distance would have converted one string into the other at an even smaller cost. Thus, the Hamming distance of an equal-length string pair cannot be strictly smaller than its edit distance.

## 11. Comprehensive GPU Scaling Results

We explore the scaling of Scrooge's throughput with GPU threads for W=64 and O=33. Supplementary Figure 11 is the extended version of Figure 8, showing all possible combinations of improvements. In particular, it shows combinations including and excluding Early Termination (ET). From Supplementary Figure 11, we observe that ET yields a small but consistent benefit when enabled. In contrast, enabling each of the memory improvements (i.e., SENE and DENT) yields a large throughput increase. Thus, we omit ET from Figure 8 for readability.

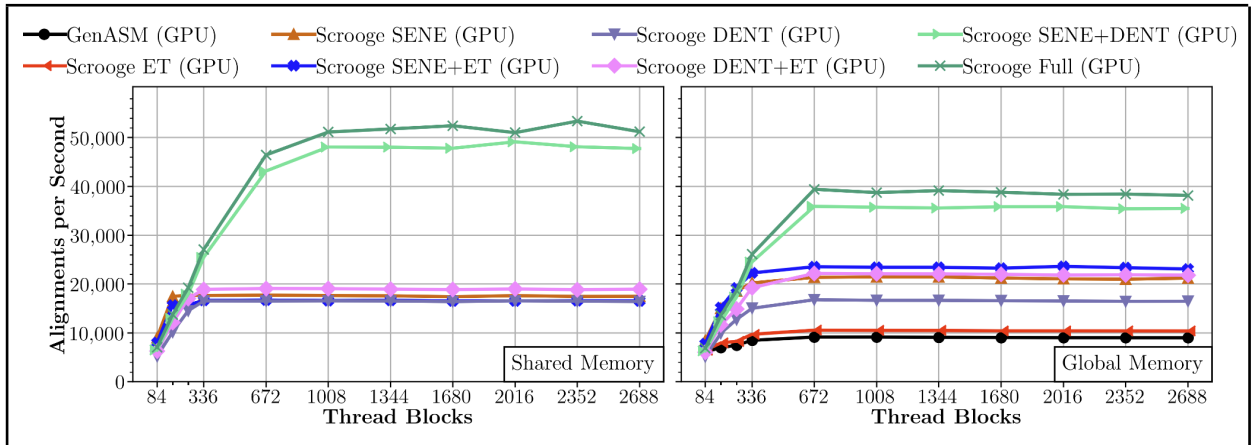

Suppl. Fig. 11: Scaling experiments of our GPU implementation with W=64, O=33, when the DP table placed in shared memory (left) and global memory (right).

## 12. Comprehensive CPU Scaling and Sensitivity to W Results

We explore the scaling of Scrooge's throughput with CPU threads for W=64 and O=33, and the sensitivity of throughput to the window size (W). Supplementary Figure 12 is the extended version of Figure 9, showing all possible combinations of improvements. In particular, it shows combinations including and excluding DENT. From Supplementary Figure 12, we observe that DENT yields small benefits at best (e.g., SENE+DENT vs.SENE), but can even lead to a slowdown (e.g., GenASM vs. DENT, SENE+ET vs. Full). In particular, for W=64 and O=33, the fastest configuration is SENE+ET doesn't use DENT. In contrast, enabling SENE and DENT yields consistent throughput increases. Thus, we omit DENT from Figure 9 for readability.

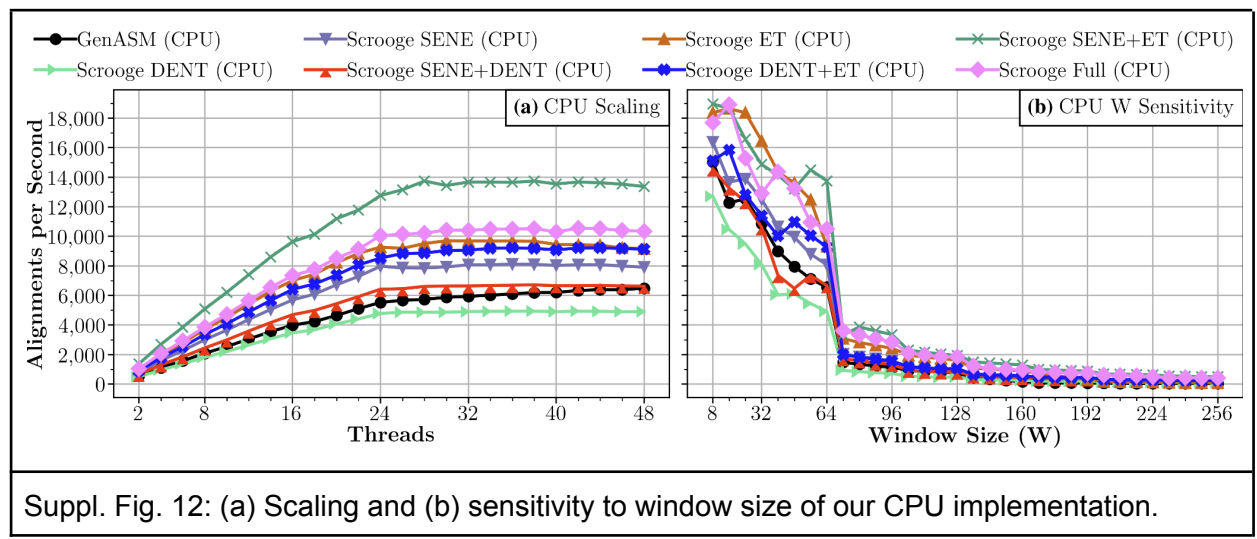

Suppl. Fig. 12: (a) Scaling and (b) sensitivity to window size of our CPU implementation.

## 13. Example of a "Difficult" Sequence Pair

We show in Section 3.6 that larger windowing parameters W and O generally improve alignment accuracy and that most sequence pairs are aligned optimally or near optimally. By manually inspecting a few worst-case (i.e., poorly aligned) sequence pairs, we observe that their apparent "difficulty" comes from small regions that (1) are relatively noisy and/or (2) contain repetitions of sequence pairs. We observe that if the windowing parameter W is increased to be larger than the noisy and/or repetitive region, these difficult sequence pairs *are* aligned correctly.

To further illustrate this observation, we give an example of an apparently "difficult" sequence pair from the long read groundtruth dataset in Supplementary Figure 13. Supplementary Figure 13 shows the alignment path produced by Edlib (i.e., a globally optimal edit distance alignment) and alignments produced by Scrooge for the windowing parameters (W=32, O=17) and (W=16, O=9). We observe that with (W=32, O=17), Scrooge aligns the sequence pair near optimally (the light green alignment by Scrooge almost completely covers the red alignment by Edlib). In contrast, Scrooge with (W=16, O=9) fails to find part of the optimal alignment early in the sequence pair (in the top left of Supplementary Figure 13) and never recovers for the remainder of the sequence pair.

Supplementary Figure 14 shows a zoomed-in view of where (W=16, O=9) first diverges from the optimal alignment. We make two key observations about this region of the sequence pair. First, we observe that it is particularly noisy, i.e., the optimal alignment contains 5 edits between reference characters 26 to 40 (an error rate of 33%). Second, we observe that it contains the highly repetitive strings "AGAGA" and "ACACACA" between reference characters 25 and 35. Scrooge with (W=16, O=9) fails to find the globally optimal alignment of these repetitive strings in favor of a more locally optimal alignment (i.e., it incurs only 2 edits until read character 31 instead of 3, which would be globally optimal). Scrooge with (W=32, O=17) also aligns "AGAGA" wrongly between reference characters 25 and 29 but has a sufficiently global optimal view to recover to the optimal alignment afterward. We conclude that the larger W parameter enables Scrooge with (W=32, O=17) to have a more globally optimal view of the noisy and repetitive region and hence produces a significantly better alignment.

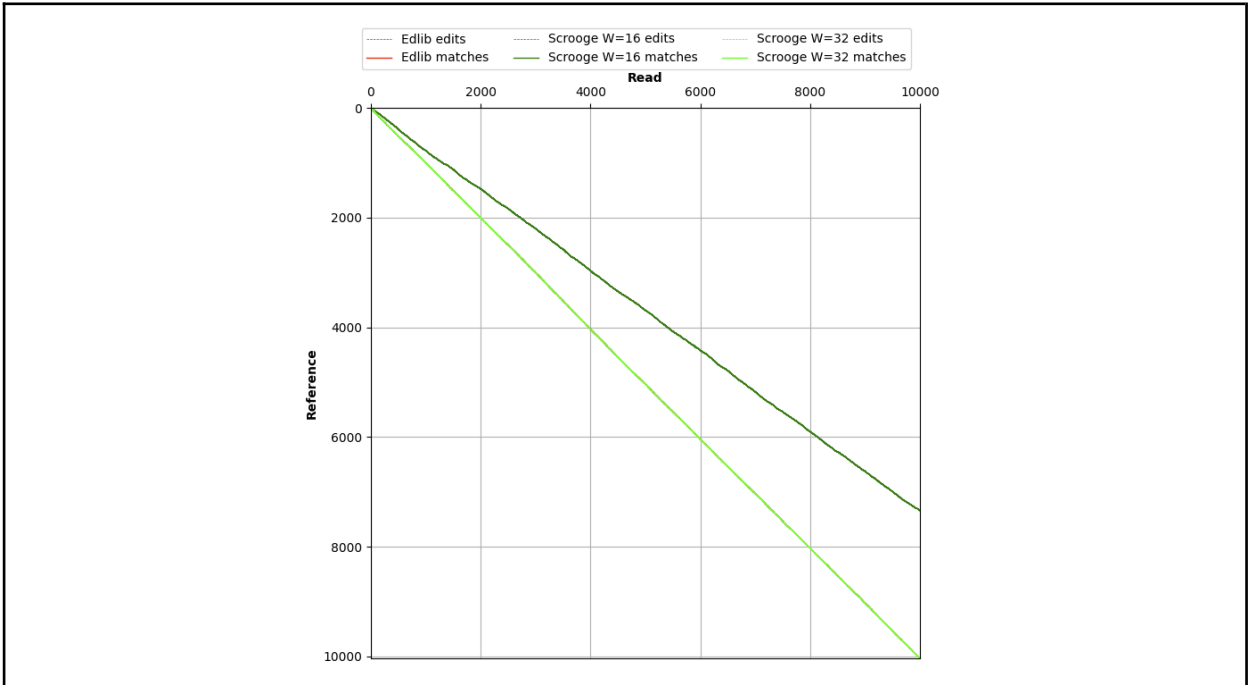

Suppl. Fig. 13: Alignment paths generated by Edlib and Scrooge with (W=32, O=17) and (W=16, O=9) for a single "difficult" sequence pair from the long read groundtruth dataset.

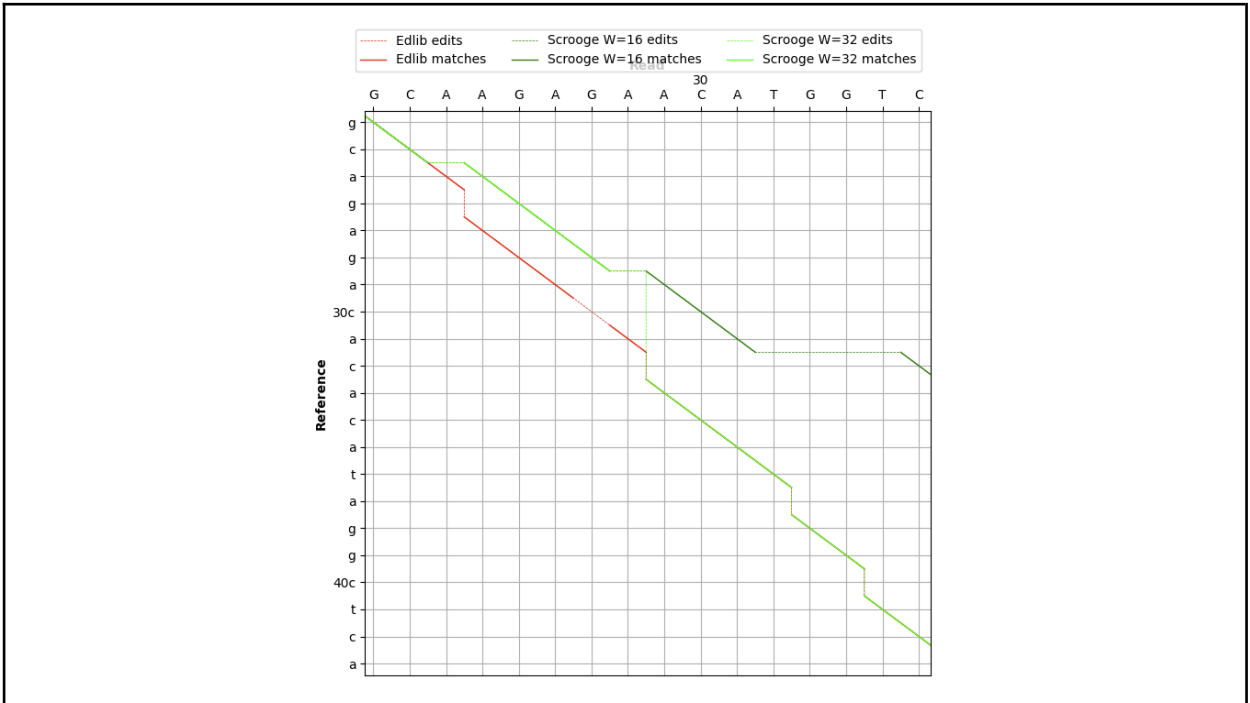

Suppl. Fig. 14: A zoomed-in view of the top left of Supplementary Figure 13, where Scrooge first diverges from the globally optimal alignment.

## 14. BiWFA Throughput Results

We repeated our throughput evaluation to include BiWFA [8] by running WFA2-lib [9] in ultralow memory mode with the edit distance and affine gap cost metrics. Supplementary Figure 15 shows the results. For the long read dataset, we observe a 36.1x and 10.4x speedup over BiWFA with the affine gap and edit distance cost metric, respectively. For the short read dataset, we observe a 4.6x and 1.1x speedup over BiWFA with the affine gap and edit distance cost metric, respectively.

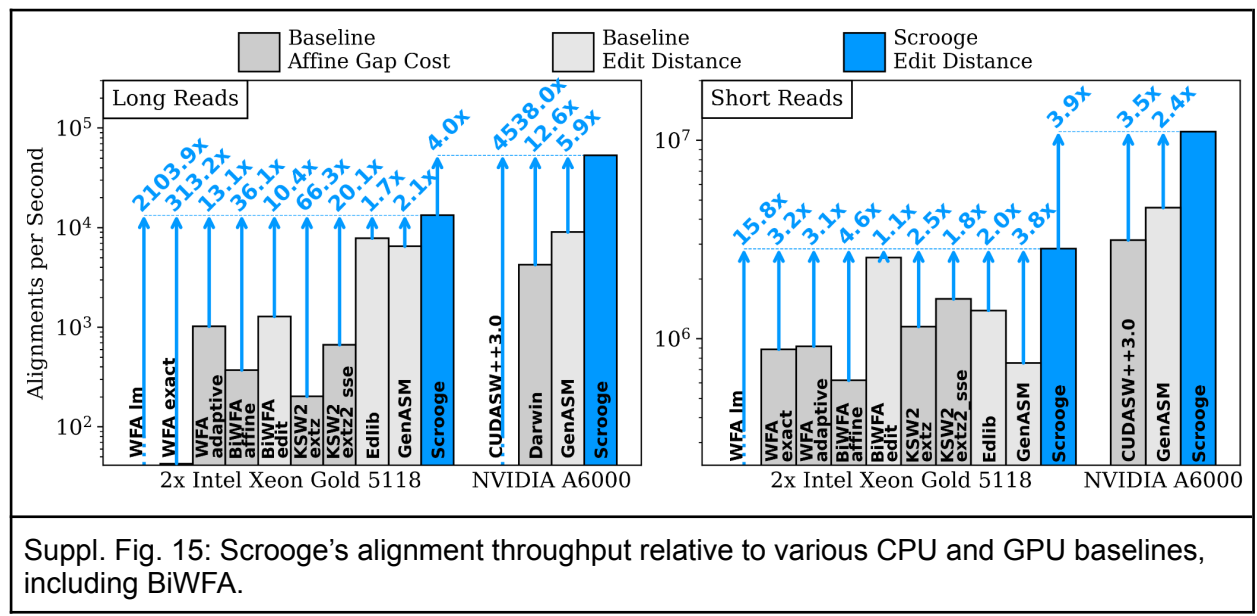

Suppl. Fig. 15: Scrooge's alignment throughput relative to various CPU and GPU baselines, including BiWFA.



\end{document}